\documentclass[prd,amsfonts,aps,nofootinbib,notitlepage,11pt,superscriptaddress]{revtex4-1}

\pdfoutput=1
\usepackage{amsmath,amssymb}
\usepackage{graphicx}
\usepackage[utf8]{inputenc}
\usepackage{cancel}
\usepackage[colorlinks=true]{hyperref}
\usepackage{color}
\usepackage{array,multirow} 
\usepackage{subcaption}

\newcommand{\UPTC}{Escuela de Física, Universidad Pedagógica y Tecnológica de Colombia,\\
Avenida Central del Norte \# 39-115, Tunja, Colombia}
\newcommand{\UdeA}{Instituto de Física, Universidad de Antioquia,\\Calle 70 \# 52-21, Apartado Aéreo 1226, Medellín, Colombia}
\newcommand{\ICTP}{Abdus Salam International Centre for Theoretical Physics,\\ Strada Costiera 11, 34151, Trieste, Italy.}

\begin{document}
\title{Multi-component scalar dark matter from  a $Z_N$ symmetry:\\ a systematic analysis}

\author{Carlos E. Yaguna}\email{carlos.yaguna@uptc.edu.co}
\affiliation{\UPTC}
\author{Óscar Zapata}\email{oalberto.zapata@udea.edu.co}
\affiliation{\UdeA}\affiliation{\ICTP}
%\date{\today}
          
\begin{abstract}
The dark matter may consist not of one elementary particle but of different species, each of them contributing a fraction of the observed dark matter density. A major theoretical difficulty with this scenario --dubbed multi-component dark matter-- is to explain the stability of these distinct particles. Imposing a single $Z_N$ symmetry, which may be a remnant of a spontaneously broken $U(1)$ gauge symmetry, seems to be the simplest way to simultaneously stabilize several dark matter particles. In this paper we systematically study scenarios for multi-component dark matter based on  various  $Z_N$ symmetries ($N\leq 10$) and with different sets of scalar fields charged under it. A generic feature of these scenarios is that the number of stable particles is not determined by the Lagrangian but depends on the relations among the masses of the different fields charged under the $Z_N$ symmetry. We explicitly obtain and illustrate the regions of parameter space that are consistent with up to five dark matter particles. For $N$ odd, all these particles turn out to be complex, whereas for $N$ even one of them may be real.  Within this  framework, many new models for multi-component dark matter can be implemented.
\end{abstract}

\maketitle
\section{Introduction}
The fundamental nature of the dark matter remains one of the most important open problems in particle and astro-particle physics. It is often assumed that the observed dark matter density, which amounts to about $25\%$ of the energy budget of the Universe~\cite{Aghanim:2018eyx}, is entirely explained by one new elementary particle --a neutralino, an axion, a new fundamental scalar, or any of the numerous other candidates that have been considered in the literature \cite{Feng:2010gw}. It may also be, though, that the dark matter is actually composed of several species, each of them contributing just a fraction of the observed dark matter density~\cite{Boehm:2003ha,Ma:2006uv,Cao:2007fy,Hur:2007ur,Lee:2008pc,Zurek:2008qg,Profumo:2009tb,Baer:2011hx}. These \emph{multi-component} dark matter scenarios have not received as much attention but they are entirely compatible with current observations (see for instance~\cite{Aoki:2012ub,Belanger:2014vza,Esch:2014jpa,Rodejohann:2015lca,Arcadi:2016kmk,Ahmed:2017dbb}). 
%or even several species such as WIMPs and axions \cite{Baer:2011hx,Bae:2013hma,Dasgupta:2013cwa,Alves:2016bib,Ma:2017zyb,Carvajal:2018ohk})

From the theoretical point of view, models with multi-component dark matter typically suffer from a crucial difficulty: the explanation of the stability of the different particles that make up the dark matter. In fact, this is a problem even for models  with just one dark matter particle. We still do not understand why this new particle is cosmologically stable. In the standard WIMP paradigm, for instance, the  dark matter particle is expected to be heavier than all or most of the known particles, which renders its stability rather puzzling.  The most common approach to stabilize the dark matter particle is to make it odd under a new $Z_2$ symmetry while the SM fields are assumed to be even. To  stabilize two or more dark matter particles, several $Z_2$'s might be used  (e.g. $Z_2\otimes Z_2'$)  but these constructions become rather awkward and difficult to implement within gauge extensions of the SM. A more appealing alternative for multi-component dark matter is to use a single $Z_N$ symmetry, with $N\geq 4$. Surprisingly, these scenarios have not been studied in detail so far \cite{Batell:2010bp,Belanger:2014bga}.

In this paper, we systematically analyze scenarios for multi-component dark matter in which the dark matter particles are scalar fields charged under a $Z_N$. Specifically, we consider extensions of  the SM  by  a number of complex scalar fields that  are SM singlets but have non-trivial $Z_N$ charges, and obtain the conditions that allow to stabilize up to five ($N\leq 10$) dark matter particles. For $N$ odd, all of them are complex fields while one of them may be real for $N$ even. In most cases, we find that  the number of dark matter (stable) particles is not determined by the Lagrangian but depends, via kinematic constraints, on the relations among the masses of the different fields. The regions of parameter space that  allow to realize multi-component dark matter scenarios are derived for each case and  illustrated  graphically in several instances. The new dark matter processes that are expected in these scenarios are also discussed. These  results should serve as a first step towards a detailed phenomenological study of  the different models for multi-component scalar dark matter that are based on  a $Z_N$ symmetry.

The rest of the paper is organized as follows. In the next section, we present the basic setup and introduce the notation we are going to follow. Our main results are then presented in sections \ref{sec:complex} and \ref{sec:real}. In them, we analyze on a case by case basis multi-component dark matter scenarios with different $Z_N$ symmetries and with varying number of fields charged under it.  In Section \ref{sec:disc}, generic features of these scenarios are briefly examined, and a couple  of possible extensions are described.  A summary of our main results is given in the conclusions whereas two special topics are, for clarity, relegated to the appendices.

\section{Framework}
The group $Z_N$ comprises the $N$ $N\mathrm{th}$ roots of $1$: $Z_N=\left\{e^{i2\pi j/N},\, j=0,1,\ldots, N-1\right\}$. Our proposal is to extend the SM with an extra $Z_N$ symmetry and few additional scalar fields that are charged under it. These extra fields would constitute the dark matter while the $Z_N$ symmetry would be responsible for stabilizing them.  Theoretically, a $Z_N$  symmetry is well motivated, for it appears as a remnant from the spontaneous breaking of either a $U(1)_X$ gauge symmetry by a scalar field $S$ with $X$ charge equal to $N$ \cite{Krauss:1988zc,Martin:1992mq} --see appendix \ref{sec:uone} for an example-- or a $SU(N)$ gauge group by a scalar multiplet transforming as the adjoint representation (recall that $Z_N$ is the center of $SU(N)$) \cite{Walker:2009en}.  Thus, dark matter stability may be closely related to gauge extensions of the SM such as  GUTs. Moreover, in this kind of setups the stability of the dark matter would automatically be protected against quantum-gravitational effects \cite{Krauss:1988zc}.

The possible charge assignments to a scalar field $\phi$ under a $Z_N$ symmetry are
\begin{align}
    1,w, w^2, ...,w^{N-1}, \,\,\,\text{with}\,\,\,w=\exp(i2\pi/N). 
\end{align}
Our goal is to find minimal setups, for different values of $N$ and with few scalar fields charged under the $Z_N$, that allow to simultaneously stabilize several particles and thus realize multi-component dark matter scenarios. Throughout, the SM particles are assumed to be singlets under this $Z_N$ symmetry. 

To begin with, the scalar fields should  have non-trivial charges (a $Z_N$ singlet would be unstable) and their charges should all be different from each other. When  two or more fields have the same charges,  they mix with one another and only the lightest one can be stable. Similarly,  the mixing terms between two different fields should be forbidden. And since $(w^\alpha)^*=w^{-\alpha}=w^{N-\alpha}$, the maximum number of  scalar fields charged under a $Z_N$ that we need to consider is $N/2$.  

We assume, therefore,  the existence of $k$ complex scalar fields $\phi_\alpha$  that are singlets of the SM gauge group and have different  $Z_N$ charges
\begin{align}
\label{eq:fields}
    \phi_\alpha\sim w^\alpha,  \,\,\,\text{with}\,\,\,\alpha=1,2,...,k,\,\,\,\text{and}\,\,\,k\leq N/2.
%    \begin{cases}
%    N/2,\,\,\,\text{for}\,\,\, N \,\,\,\text{odd},\\
%    N/2-1,\,\,\,\text{for}\,\,\, N \,\,\,\text{even}.    
%    \end{cases} 
\end{align}
We further require that these scalar fields do not develop a vacuum expectation value so that the $Z_N$ symmetry remains unbroken. Notice, in particular, that  the scenario with $k$ DM particles may be minimally realized by a $Z_{2k}$ symmetry. A $Z_4$ is, therefore, the lowest $Z_N$ symmetry consistent with multi-component dark matter.

Among the Lagrangian terms that are $Z_N$ invariant, there will usually be some that can induce the decay of one of the scalar fields into others. They correspond to cubic and quartic interactions  involving  $\phi_\alpha$ only once, and they lead to  two- and three-body decays of $\phi_\alpha$ into other $\phi_\beta$'s ($\alpha\neq \beta$).  The terms $\phi_1\phi_2^2$ and $\phi_1^3\phi_2$, for instance, are  both invariant under a $Z_5$ and   would lead to $\phi_1\to 2\phi_2$ and  $\phi_2\to 3\phi_1$ respectively. Hence, for such a $\phi_\alpha$ to  be a dark matter particle, one must ensure that all its possible decays are kinematically forbidden\footnote{A more complicated alternative is to impose extra symmetries to forbid such interactions.}, which entails  restrictions on the masses of the scalar fields. In the example mentioned above they would read $M_1<2M_2$ and $M_2<3M_1$, being $M_{1,2}$ the masses of $\phi_{1,2}$. The number of stable (dark matter) particles is thus not determined by the Lagrangian itself but depends, due to kinematic constraints, on the relations among the masses of the different fields present in the model. As we will see in the next two sections, this is a generic feature of multi-component dark matter scenarios with a $Z_N$ stabilizing symmetry. It becomes necessary, therefore, to determine, on a case-by-case basis, the regions of parameter space that realize multi-component dark matter.

\section{Stability analysis: Complex Dark Matter}
\label{sec:complex}

We next investigate the possible realizations of  multi-component dark matter scenarios for different $Z_N$ symmetries and several sets of scalar fields $\phi_\alpha$  charged under it. In this section, we restrict ourselves to $\alpha<N/2$ while the cases with  $\alpha=N/2$ will be examined in the next section. Since  $\alpha\neq N/2$, the only quadratic terms allowed by the $Z_N$ are of the form $\phi_\alpha^\dagger\phi_\alpha$, and thus  the fields $\phi_\alpha$ are themselves  mass eigenstates. Consequently,  the dark matter particles will  all be complex scalar fields.   

Specifically, we present  in this section all the  operators of mass dimension $3$ and $4$ ($d=3,4$) that are allowed by the $Z_N$ symmetry and use them to determine the  regions of parameter space compatible with  multi-component dark matter. These regions are then illustrated graphically for the most relevant cases.  For completeness, we also include  the non-renormalizable $d=5$ terms that may induce new decays of one dark matter particle into others. Let us emphasize, though, that, due to the absence of visible particles in the final states, these processes are not constrained by indirect detection searches.

In this section, the notation $\phi_j^\dagger\to\phi_{sj}$ will be used, and the mass of the complex field $\phi_j$ will be denoted by $M_j$. Notice, in particular, that $M_1$ does not denote the mass of the lightest  scalar field but rather the mass of $\phi_1\sim w$. In the following, we  analyze, one by one, the different $Z_N$ symmetries, for $N\leq 10$, that lead to  multi-component dark matter.

\subsection{$Z_5$}
\begin{figure}[t]
\centering
\includegraphics[scale=0.5]{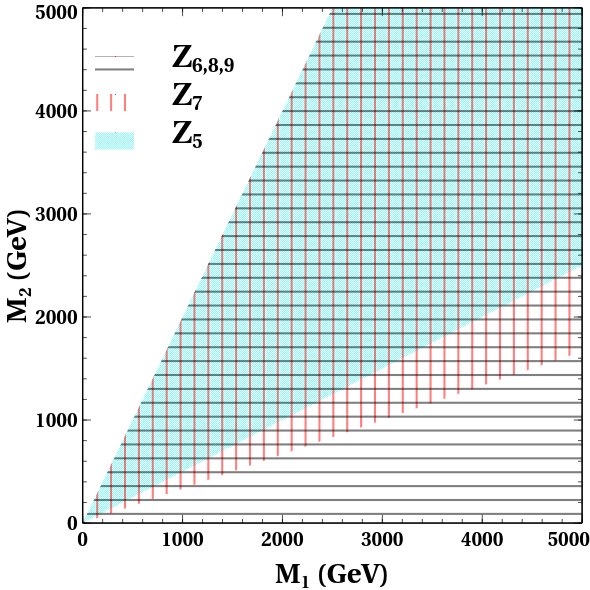}\includegraphics[scale=0.5]{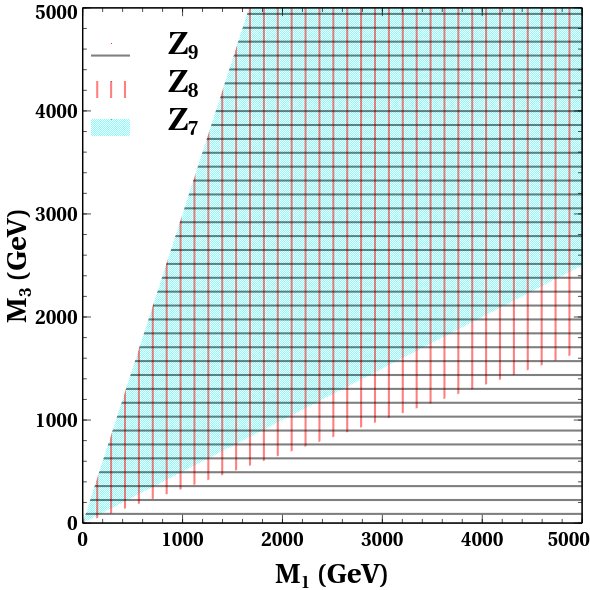}
\caption{Left (right): regions in the plane $(M_1,M_2)$ where $\phi_1$ and $\phi_2$ (in the plane $(M_1,M_3)$ where $\phi_1$ and $\phi_3$) are both simultaneously stable  for different $Z_N$ symmetries.}
\label{fig:twoparticlesp1p2}
\end{figure}

$Z_5$ is the lowest $Z_N$ that allows to realize a two-component dark matter scenario where both particles are complex scalar fields. The charge assignment of the two fields is uniquely determined to be $(\phi_1,\phi_2)$.  The invariant interaction terms are given by
\begin{align}
 \mathcal{L}_3\supset&\,\, \phi _1 \phi _2^2,\phi _2 \phi _{\text{s1}}^2.\\
  \mathcal{L}_4\supset&\,\, \phi _1^3 \phi _2,\phi _1^2 \phi _{\text{s1}}^2,\phi _1 \phi _2 \phi _{\text{s1}} \phi
   _{\text{s2}},\phi _2^2 \phi _{\text{s2}}^2,\phi _1 \phi _{\text{s2}}^3 .%\\
   %\mathcal{L}_5\supset&\,\, 0.
 \end{align}
Accordingly, each field may have two- and three-body decays into the other.  Their simultaneous stability is reached for $\frac{M_1}{2}<M_2<2 M_1$ --see the shaded green region in  the left panel of Fig.~\ref{fig:twoparticlesp1p2}. In this case,   the $d=5$ non-renormalizable decay operators are forbidden due to the $Z_5$ charge assignment. 

\subsection{$Z_6$}
We can only have two different fields charged under the symmetry: $\phi_1$, $\phi_2$. 
The corresponding interaction terms are
\begin{align}
 \mathcal{L}_3\supset&\,\, \phi _2^3,\phi _1^2 \phi _{\text{s2}}.\\
  \mathcal{L}_4\supset&\,\, \phi _1^2 \phi _2^2,\phi _1^2 \phi _{\text{s1}}^2,\phi _1 \phi _2 \phi _{\text{s1}} \phi _{\text{s2}},\phi _2^2 \phi
   _{\text{s2}}^2.\\
   \mathcal{L}_5\supset&\,\, \phi _1^4 \phi _2.
 \end{align}
 Thus, $\phi_1$ is automatically stable while $\phi_2$ will be stable for $M_2<2M_1$, which in turn implies that the four-body decay  $\phi_2\to 4\phi_1$ induced by $d=5$ operators is kinematically closed. The region in the plane $(M_1,M_2)$ where a two-component dark matter scenario is obtained is represented by horizontal grid 
 lines in the left panel of Fig.~\ref{fig:twoparticlesp1p2}.

\subsection{$Z_7$}
In this case we can have up to three fields charged under the symmetry: $\phi_1$, $\phi_2$, $\phi_3$.

\subsubsection{Two fields}
\begin{itemize}
    \item ($\phi_1,\phi_2$): The interaction terms are
\begin{align}
 \mathcal{L}_3\supset&\,\,\phi _1^2 \phi _{\text{s2}}.\\
  \mathcal{L}_4\supset&\,\, \phi _1 \phi _2^3,\phi _1^2 \phi _{\text{s1}}^2,\phi _1 \phi _2 \phi _{\text{s1}} \phi _{\text{s2}},\phi _2^2 \phi
   _{\text{s2}}^2. \\
   \mathcal{L}_5\supset&\,\,\phi _1 \phi _{\text{s2}}^4.
 \end{align}

 The two-body decay $\phi_2\to2\phi_1$ and the three-body decay $\phi_1\to3\phi_2$ can be  forbidden through the condition $\frac{M_2}{2}<M_1<3 M_2$, which also ensures that the four-body decay of $\phi_1$ via $d=5$ operators is  kinematically closed. The mass constraint leads to the viable region (for two-component dark matter) represented by vertical grid lines in the left panel of Fig.~\ref{fig:twoparticlesp1p2}. 
 
    \item ($\phi_1,\phi_3$): The interaction terms are
\begin{align}
 \mathcal{L}_3\supset&\,\,  \phi _1 \phi _3^2.\\
  \mathcal{L}_4\supset&\,\, \phi _1^2 \phi _{\text{s1}}^2,\phi _1^3 \phi _{\text{s3}},\phi _1 \phi _3 \phi _{\text{s1}}
   \phi _{\text{s3}},\phi _3^2 \phi _{\text{s3}}^2 . \\
   \mathcal{L}_5\supset&\,\, \phi _1^4 \phi _3.
 \end{align}
 The stability of both fields is thus ensured by the condition $\frac{M_1}{2}<M_3<3 M_1$ --see the  horizontal grid lines in the right panel of Fig.~\ref{fig:twoparticlesp1p2}. Notice that this condition automatically guarantees that the $d=5$ operators do not induce any decay. 
 
    \item ($\phi_2,\phi_3$): The interaction terms are
\begin{align}
 \mathcal{L}_3\supset&\,\,  \phi _2^2 \phi _3.\\
  \mathcal{L}_4\supset&\,\, \phi _2^2 \phi _{\text{s2}}^2,\phi _2 \phi _3 \phi _{\text{s2}} \phi _{\text{s3}},\phi _3^2
   \phi _{\text{s3}}^2,\phi _2 \phi _{\text{s3}}^3 . \\
   \mathcal{L}_5\supset&\,\, \phi _2 \phi _3^4.
 \end{align}
It follows that the stability condition is $\frac{M_3}{2}<M_2<3 M_3$, which automatically prevents decays via $d=5$ operators.  
\end{itemize}

\subsubsection{Three fields}
The interaction terms for the $(\phi_1,\phi_2,\phi_3)$ scenario are given by
\begin{align}
 \mathcal{L}_3\supset&\,\,  \phi _2^2 \phi _3,\phi _1 \phi _3^2,\phi _1^2 \phi _{\text{s2}},\phi _1 \phi _2 \phi _{\text{s3}}.\\
  \mathcal{L}_4\supset&\,\, \phi _1 \phi _2^3,\phi _1^2 \phi _2 \phi _3,\phi _2 \phi _3^2 \phi _{\text{s1}},\phi _1^2 \phi _{\text{s1}}^2,\phi _3
   \phi _{\text{s1}}^3,\phi _3^3 \phi _{\text{s2}},\phi _1 \phi _2 \phi _{\text{s1}} \phi _{\text{s2}},\phi _2^2 \phi
   _{\text{s2}}^2,\phi _1 \phi _3 \phi _{\text{s2}}^2,\nonumber\\
   &\phi _1 \phi _3 \phi _{\text{s1}} \phi _{\text{s3}},\phi _2 \phi _3 \phi
   _{\text{s2}} \phi _{\text{s3}},\phi _3^2 \phi _{\text{s3}}^2. \\
   \mathcal{L}_5\supset&\,\, \phi _1^4 \phi _3,\phi _2 \phi _3^4,\phi _1 \phi
   _{\text{s2}}^4.
 \end{align}
 
\begin{figure}
\centering
\includegraphics[scale=0.5]{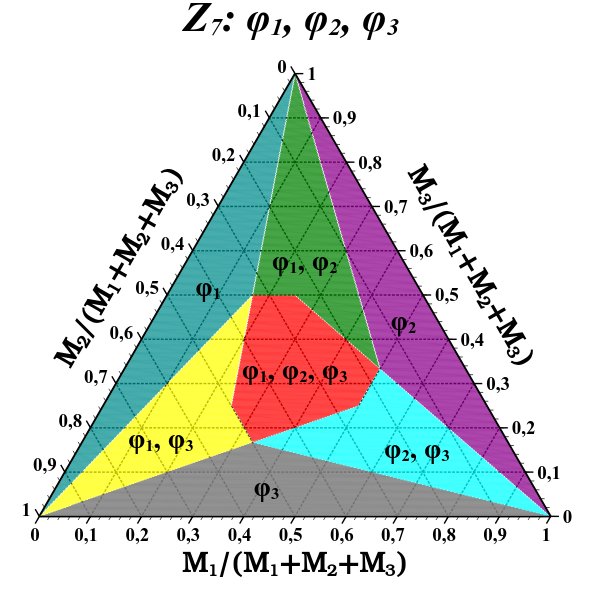}\includegraphics[scale=0.5]{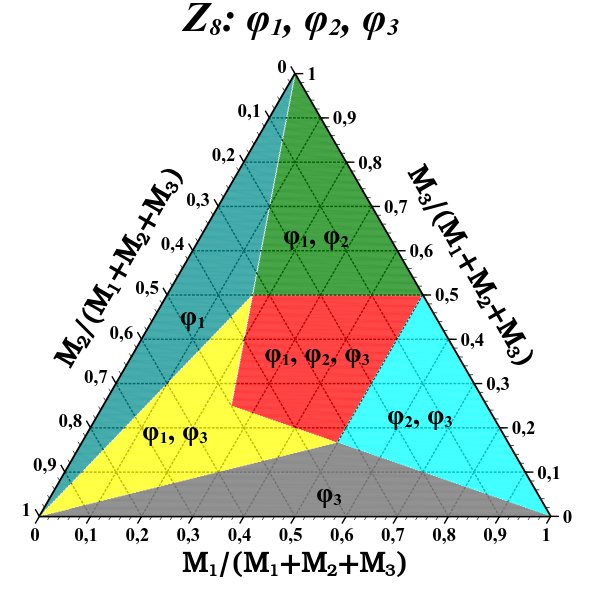}\\
\vspace{0.5cm}
\includegraphics[scale=0.5]{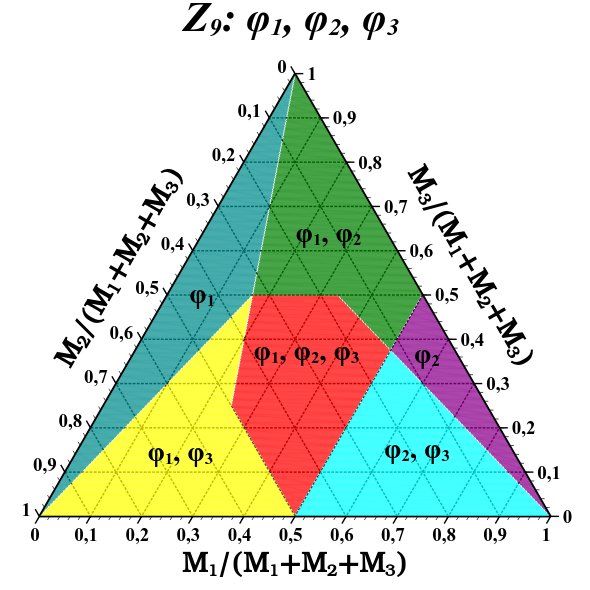}\includegraphics[scale=0.5]{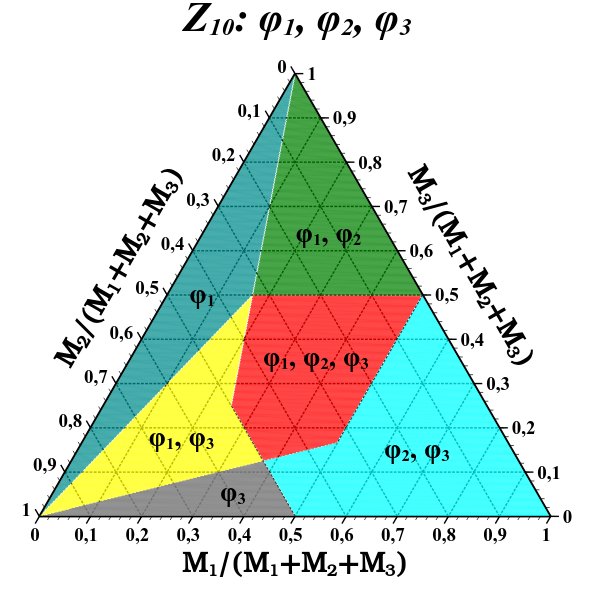}\caption{Stability regions for a $Z_7$ (top left panel),  $Z_8$ (top right panel), $Z_9$ (bottom left panel) and $Z_{10}$ (bottom right panel) symmetry with fields  $\phi_1,\phi_2,\phi_3$.}
%\label{fig:trianglez7}
\label{fig:trianglep123}
\end{figure}

Hence,  all the three fields have two- and three-body decays.  They  can however  be forbidden by imposing
$M_2<2 M_1,\, M_3<2 M_2,\, M_1<2 M_3,\, M_1<M_2+M_3,\, M_2<M_1+M_3,\,M_3<M_1+M_2$.
%%%%$M_1<2M_3,3M_2,M_2+M_3$ $\land$ $M_2<2M_1,3M_3,M_1+M_3$ $\land$ $M_3<3M_1,2M_3,M_1+M_2$.
%$\left(\frac{M_1}{3}<M_2\leq \frac{M_1}{2}\land M_1-M_2<M_3<2M_2\right)\,\,\lor$
%$\left(\frac{M_1}{2}<M_2\leq M_1\land \frac{M_1}{2}<M_3<2 M_2\right)\,\,\lor$
%$\left(M_1<M_2\leq \frac{3 M_1}{2}\land \frac{M_1}{2}<M_3<M_1+M_2\right)\,\,\lor$ 
%$\left(\frac{3 M_1}{2}<M_2<2 M_1\land M_2-M_1<M_3<M_1+M_2\right)$.
The different stability regions are illustrated in the top left panel of Fig.~\ref{fig:trianglep123}, which is  a ternary diagram with normalized axis $M_i/(M_1+M_2+M_3)$. In the central (red) region all  three fields ($\phi_1, \phi_2, \phi_3$) are stable. As we will see, this is a common feature of these scenarios with multi-component dark matter: stability for all the fields is usually achieved in the region of the parameter space where the masses are not that different from each other (in the central part of a ternary plot). This does not mean, though, that the masses have to  be degenerate. One can see from the figure, for example, that the point $(0.2, 0.35, 0.45)$ lies inside the central (red) region. Thus, all three particles are stable for $M_1=200$ GeV, $M_2=350$ GeV, and  $M_3=450$ GeV, which is not a degenerate or compressed spectrum. 

In the same figure, we also display the stability regions for two (and one) dark matter particles.  The $Z_7$ symmetry with fields $(\phi_1,\phi_2,\phi_3)$ is rather special because  the seven possible cases are all realized in certain regions of parameter space.

\subsection{$Z_8$}
The maximum number of fields charged under this symmetry is again three: $\phi_1$, $\phi_2$ and $\phi_3$.
\subsubsection{Two fields}
\begin{itemize}
    \item $(\phi_1,\phi_2)$:
    The interaction terms are
\begin{align}
 \mathcal{L}_3\supset&\,\, \phi _1^2 \phi _{\text{s2}}.\\
  \mathcal{L}_4\supset&\,\, \phi _2^4,\phi _1^2 \phi _{\text{s1}}^2,\phi _1 \phi _2 \phi _{\text{s1}} \phi _{\text{s2}},\phi _2^2 \phi
   _{\text{s2}}^2.
%  \mathcal{L}_5\supset&\,\, 0.
 \end{align}
  It follows that $\phi_1$ is always stable whereas $\phi_2$ becomes stable for $M_2<2M_1$, and there are no further decays via $d=5$ operators. The viable region in the plane $(M_1,M_2)$ is shown in the left panel of Fig.~\ref{fig:twoparticlesp1p2}.  
\item $(\phi_1,\phi_3)$:   In this case there are only  quartic interactions: 
\begin{align}
% \mathcal{L}_3\supset&\,\, 0 .\\
  \mathcal{L}_4\supset&\,\, \phi _1^2 \phi _3^2,\phi _3^3 \phi _{\text{s1}},\phi _1^2 \phi _{\text{s1}}^2,\phi _1^3 \phi
   _{\text{s3}},\phi _1 \phi _3 \phi _{\text{s1}} \phi _{\text{s3}},\phi _3^2 \phi _{\text{s3}}^2.
%  \mathcal{L}_5\supset&\,\, 0 .
 \end{align}
They can induce the three-body decay of the heavier into the lighter. 
Consequently,  the stability condition for both particles reads $\frac{M_1}{3}<M_3<3 M_1$ --see right panel of Fig.~\ref{fig:twoparticlesp1p2}.
\item $(\phi_2,\phi_3)$:   The interaction terms are:
\begin{align}
 \mathcal{L}_3\supset&\,\, \phi _2 \phi _3^2.\\
  \mathcal{L}_4\supset&\,\, \phi _2^4,\phi _2^2 \phi _{\text{s2}}^2,\phi _2 \phi _3 \phi _{\text{s2}} \phi _{\text{s3}},\phi _3^2
   \phi _{\text{s3}}^2.
%  \mathcal{L}_5\supset&\,\, 0 .
 \end{align}
In this case $\phi_3$ is always stable whereas the  two-body decay of $\phi_2$ is closed as long as $M_2<2M_3$.  There are no $d=5$ operators inducing the decay of either field.  
\end{itemize}
\subsubsection{Three fields}
The only choice is   $(\phi_1,\phi_2,\phi_3)$ with the following interaction terms:
\begin{align}\label{eq:Z8phi123-L3}
 \mathcal{L}_3\supset&\,\, \phi _2 \phi _3^2,\phi _1^2 \phi _{\text{s2}},\phi _1 \phi _2 \phi _{\text{s3}}.\\
  \mathcal{L}_4\supset&\,\, \phi _2^4,\phi _1 \phi _2^2 \phi _3,\phi _1^2 \phi _3^2,\phi _3^3 \phi _{\text{s1}},\phi _1^2 \phi _{\text{s1}}^2,\phi _3
   \phi _{\text{s1}}^3,\phi _1 \phi _2 \phi _{\text{s1}} \phi _{\text{s2}},\phi _2^2 \phi _{\text{s2}}^2,\phi _1 \phi _3 \phi
   _{\text{s2}}^2,\nonumber\\
   &\phi _1 \phi _3
   \phi _{\text{s1}} \phi _{\text{s3}},\phi _2 \phi _3 \phi _{\text{s2}} \phi _{\text{s3}},\phi _3^2 \phi _{\text{s3}}^2.\label{eq:Z8phi123-L4}\\
  \mathcal{L}_5\supset&\,\,  \phi _1^3 \phi _2 \phi _3,\phi _2^3 \phi _3 \phi _{\text{s1}},\phi _1 \phi _3^3
   \phi _{\text{s2}}.\label{eq:Z8phi123-L5}
 \end{align}
Accordingly,  each $\phi_i$ potentially has two-body decays while only $\phi_1$ and $\phi_3$ have additional three-body decays. The full stability regions are shown in the top right panel of Fig.~\ref{fig:trianglep123}. The three fields will be stable (red region) for  
$M_2<2 M_1,\, M_2<2 M_3,\, M_1<M_2+M_3,\, M_3<M_1+M_2$. 
%$\left(\frac{M_1}{3}<M_3\leq M_1\land M_1-M_3<M_2<2 M_3\right)\,\,\lor$ \\
%$\left(M_1<M_3<3 M_1\land M_3-M_1<M_2<2 M_1\right)$.  
Notice from the figure that there are just six regions in this case, for it is not possible to get $\phi_2$ as the only stable particle.

\subsection{$Z_9$}
It is possible to have up to four fields ($\phi_{1,\ldots, 4}$) charged under a $Z_9$. 

\subsubsection{Two fields}
With two fields, there are six different scenarios, which we next examine one by one.
\begin{itemize}
\item $(\phi_1,\phi_2)$: The interaction terms are
\begin{align}
 \mathcal{L}_3\supset&\,\, \phi _1^2 \phi _{\text{s2}}.\\
  \mathcal{L}_4\supset&\,\, \phi _1^2 \phi _{\text{s1}}^2,\phi _1 \phi _2 \phi _{\text{s1}} \phi
   _{\text{s2}},\phi _2^2 \phi _{\text{s2}}^2.\\
   \mathcal{L}_5\supset&\,\,\phi _1 \phi _2^4.
 \end{align}
In this case $\phi_1$ is stable at  the renormalizable level whereas $\phi_2$ is stable for $M_2<2M_1$.  The viable region is represented by the horizontal grid lines in the left panel of Fig.~\ref{fig:twoparticlesp1p2}.  To prevent the decay of $\phi_1$ via  $d=5$ operators, we  would need to impose the additional constraint $M_1<4M_2$.  
    \item $(\phi_1,\phi_3)$:  The interaction terms are
\begin{align}
 \mathcal{L}_3\supset&\,\, \phi _3^3.\\
  \mathcal{L}_4\supset&\,\,\phi _1^2 \phi _{\text{s1}}^2,\phi _1^3 \phi
   _{\text{s3}},\phi _1 \phi _3 \phi _{\text{s1}} \phi _{\text{s3}},\phi _3^2 \phi
   _{\text{s3}}^2.
%   \mathcal{L}_5\supset&\,\,0 .
 \end{align}
The unique possible decay in this case is $\phi_3\to3\phi_1$, which does not take place for $M_3<3M_1$. The  viable region is shown, as horizontal grid lines,  in the right panel of Fig.~\ref{fig:twoparticlesp1p2}. There are no $d=5$ operators inducing further decays.
\item $(\phi_1,\phi_4)$:  The interaction terms are
\begin{align}
 \mathcal{L}_3\supset&\,\,\phi _1 \phi _4^2.\\
  \mathcal{L}_4\supset&\,\, \phi _1^2 \phi _{\text{s1}}^2,\phi _1 \phi _4 \phi _{\text{s1}} \phi
   _{\text{s4}},\phi _4^2 \phi _{\text{s4}}^2.\\
   \mathcal{L}_5\supset&\,\, \phi _1^4 \phi _{\text{s4}}.
 \end{align}
The two-body decay  $\phi_1\to2\phi_4$ implies the condition $M_1<2M_4$ while the four-body decay via $d=5$ operators does not arise for  $M_4<4M_1$. 
    \item $(\phi_2,\phi_3)$:  The interaction terms are
\begin{align}
 \mathcal{L}_3\supset&\,\, \phi _3^3.\\
  \mathcal{L}_4\supset&\,\, \phi _2^3 \phi _3,\phi _2^2 \phi _{\text{s2}}^2,\phi _2 \phi _3 \phi _{\text{s2}}
   \phi _{\text{s3}},\phi _3^2 \phi
   _{\text{s3}}^2.
%   \mathcal{L}_5\supset&\,\,0 .
 \end{align}
Hence, $\phi_2$ is always stable whereas $\phi_3$ may decay via $\phi_3\to3\phi_2$. To get a two-component dark matter scenario, $M_3<3M_2$ is required.  
    \item $(\phi_2,\phi_4)$:  The interaction terms are
\begin{align}
 \mathcal{L}_3\supset&\,\, \phi _2^2 \phi _{\text{s4}}.\\
  \mathcal{L}_4\supset&\,\, \phi _2^2 \phi _{\text{s2}}^2,\phi _2 \phi _4 \phi _{\text{s2}} \phi
   _{\text{s4}},\phi _4^2 \phi _{\text{s4}}^2.\\
   \mathcal{L}_5\supset&\,\, \phi _2 \phi _4^4.
 \end{align}
As a result, $\phi_2$ is stable at the renormalizable level whereas $\phi_4$  will be stable for  $M_4<2M_2$. The decay of $\phi_2$ via  $d=5$ operators can be prevented for $M_2<4M_4$. 
\item $(\phi_3,\phi_4)$:  The interaction terms are
\begin{align}
\mathcal{L}_3\supset&\,\, \phi _3^3.\\
\mathcal{L}_4\supset&\,\, \phi _4^3 \phi _{\text{s3}},\phi _3^2 \phi _{\text{s3}}^2,\phi _3 \phi _4 \phi
   _{\text{s3}} \phi _{\text{s4}},\phi _4^2 \phi _{\text{s4}}^2.
%   \mathcal{L}_5\supset&\,\, 0.
 \end{align}
Thus, $M_3<3M_4$ leads to a two-component dark matter scenario. 
\end{itemize}
\subsubsection{Three fields}
There are four different scenarios with three fields charged under a $Z_9$: 

\begin{itemize}
    \item $(\phi_1,\phi_2,\phi_3)$:  The interaction terms are
\begin{align}
 \mathcal{L}_3\supset&\,\, \phi _3^3,\phi _1^2 \phi _{\text{s2}},\phi _1 \phi _2 \phi _{\text{s3}}.\\
  \mathcal{L}_4\supset&\,\, \phi _2^3 \phi _3,\phi _1 \phi _2 \phi _3^2,\phi _1^2 \phi _{\text{s1}}^2,\phi _3
   \phi _{\text{s1}}^3,\phi _1 \phi _2 \phi _{\text{s1}} \phi _{\text{s2}},\phi _2^2
   \phi _{\text{s2}}^2,\phi _1 \phi _3 \phi _{\text{s2}}^2,\nonumber\\
   &\phi _1 \phi _3 \phi
   _{\text{s1}} \phi _{\text{s3}},\phi _2 \phi _3 \phi _{\text{s2}} \phi
   _{\text{s3}},\phi _3^2 \phi _{\text{s3}}^2.\\
   \mathcal{L}_5\supset&\,\, \phi _1 \phi _2^4,\phi _1^2 \phi _2^2 \phi _3,\phi _2^2 \phi _3^2 \phi
   _{\text{s1}}.
 \end{align}
In this case, there are potential two- and three-body decays for every field. The  stability region for all three particles is described by $M_2<2 M_1,\, M_1<M_2+M_3,\, M_2<M_1+M_3,\, M_3<M_1+M_2,\, M_3<3 M_2$
%$\left(\frac{M_1}{4}<M_2\leq \frac{M_1}{2}\land M_1-M_2<M_3<3M_2\right)\,\,\lor$ $\left(\frac{M_1}{2}<M_2\leq M_1\land M_1-M_2<M_3<M_1+M_2\right)\,\,\lor$ $\left(M_1<M_2<2 M_1\land M_2-M_1<M_3<M_1+M_2\right)$
and shown in the bottom left panel of Fig.~\ref{fig:trianglep123}  --the red region. That figure displays also the regions where a two-component dark matter scenario is realized (one of the three fields is unstable), and the regions where the standard scenario with just one dark matter particle is recovered. 

\item $(\phi_1,\phi_2,\phi_4)$:  The interaction terms are
\begin{align}
 \mathcal{L}_3\supset&\,\, \phi _1 \phi _4^2,\phi _1^2 \phi _{\text{s2}},\phi _2^2 \phi _{\text{s4}}.\\
  \mathcal{L}_4\supset&\,\, \phi _1 \phi _2^2 \phi _4,\phi _2 \phi _4^2 \phi _{\text{s1}},\phi _1^2 \phi
   _{\text{s1}}^2,\phi _1 \phi _2 \phi _{\text{s1}} \phi _{\text{s2}},\phi _4 \phi
   _{\text{s1}}^2 \phi _{\text{s2}},\phi _2^2 \phi _{\text{s2}}^2,\nonumber\\
   &\phi _1 \phi _4 \phi _{\text{s1}} \phi _{\text{s4}},\phi _2 \phi _4 \phi
   _{\text{s2}} \phi _{\text{s4}},\phi _4^2 \phi _{\text{s4}}^2.\\
   \mathcal{L}_5\supset&\,\, \phi _1 \phi _2^4,\phi _1^3 \phi _2 \phi _4,\phi _2 \phi _4^4,\phi _2^3 \phi _4
   \phi _{\text{s1}},\phi _4^3 \phi _{\text{s1}} \phi
   _{\text{s2}},\phi _1^4 \phi _{\text{s4}}.
 \end{align}

It follows that the condition $M_2<2 M_1,\, M_1<2 M_4.\, M_4<2 M_2$  
%$\frac{M_1}{4}<M_2<2 M_1\land \frac{M_1}{2}<M_4<2 M_2$
prevents all the two- and three-body decays and leads to a three-component dark matter scenario. Figure ~\ref{fig:trianglep124} displays the different stability regions for this case. Notice that all seven possibilities can be realized. Moreover, this figure is particularly symmetric: the stability region for the three particles is an equilateral triangle, the three stability regions for two particles are all the same size and shape, and the same happens with the three stability regions for a single particle.  

\begin{figure}
\centering
\includegraphics[scale=0.5]{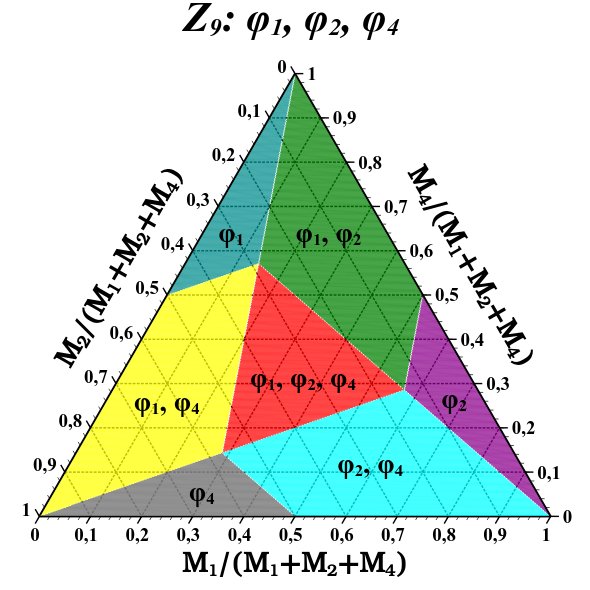}\includegraphics[scale=0.5]{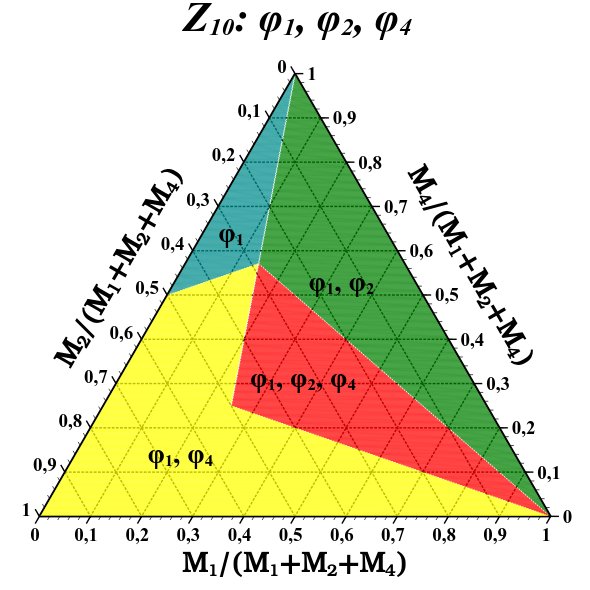}\caption{Stability regions for a $Z_9$ (left panel) and  $Z_{10}$ (right panel) symmetry with fields  $\phi_1,\phi_2,\phi_4$.}
\label{fig:trianglep124}
\end{figure}
\item $(\phi_1,\phi_3,\phi_4)$:  The interaction terms are
\begin{align}
 \mathcal{L}_3\supset&\,\, \phi _3^3,\phi _1 \phi _4^2,\phi _1 \phi _3 \phi _{\text{s4}}.\\
  \mathcal{L}_4\supset&\,\, \phi _1^2 \phi _3 \phi _4,\phi _3^2 \phi _4 \phi _{\text{s1}},\phi _1^2 \phi
   _{\text{s1}}^2,\phi _1^3 \phi _{\text{s3}},\phi _4^3 \phi
   _{\text{s3}},\phi _1 \phi _3 \phi _{\text{s1}} \phi _{\text{s3}},\phi _3^2 \phi
   _{\text{s3}}^2,\nonumber\\
   &\phi _1 \phi _4 \phi _{\text{s1}} \phi _{\text{s4}},\phi _3 \phi _4
   \phi _{\text{s3}} \phi _{\text{s4}},\phi _4^2 \phi
   _{\text{s4}}^2.\\
   \mathcal{L}_5\supset&\,\, \phi _3 \phi _4^2 \phi _{\text{s1}}^2,\phi _1^2 \phi
   _4 \phi _{\text{s3}}^2,\phi _1^4 \phi _{\text{s4}}.
 \end{align}
 
These interactions give rise to two- and three-body decays for every field. The  stability region  for the three particles is described by $M_1<2 M_4,\,M_1<M_3+M_4,\, M_3<M_1+M_4,\, M_4<M_1+M_3,\, M_3<3 M_1$ 
%$\left(0<M_3\leq \frac{M_1}{2}\land M_1-M_3<M_4<M_1+M_3\right)\,\,\lor$ $\left(\frac{M_1}{2}<M_3\leq \frac{3 M_1}{2}\land \frac{M_1}{2}<M_4<M_1+M_3\right)\,\,\lor$ $\left(\frac{3 M_1}{2}<M_3<3 M_1\land M_3-M_1<M_4<M_1+M_3\right)$
and shown in the  left panel of Fig.~\ref{fig:trianglep134}  (red region). That figure also displays the regions with one- or two-component dark matter. 

\begin{figure}
\centering
\includegraphics[scale=0.5]{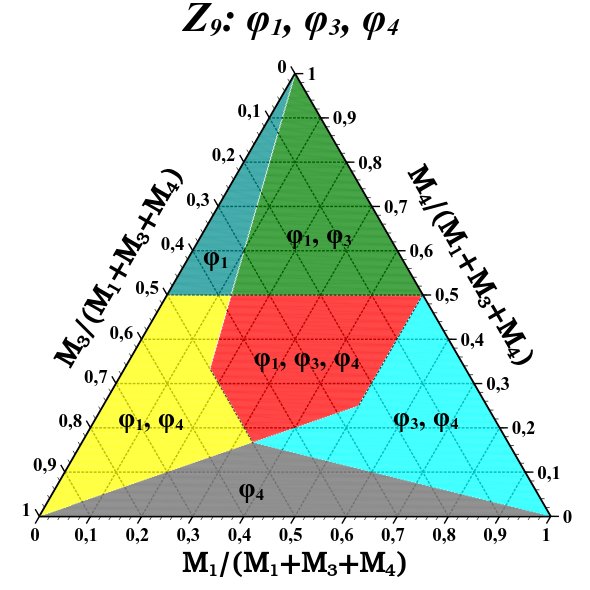}\includegraphics[scale=0.5]{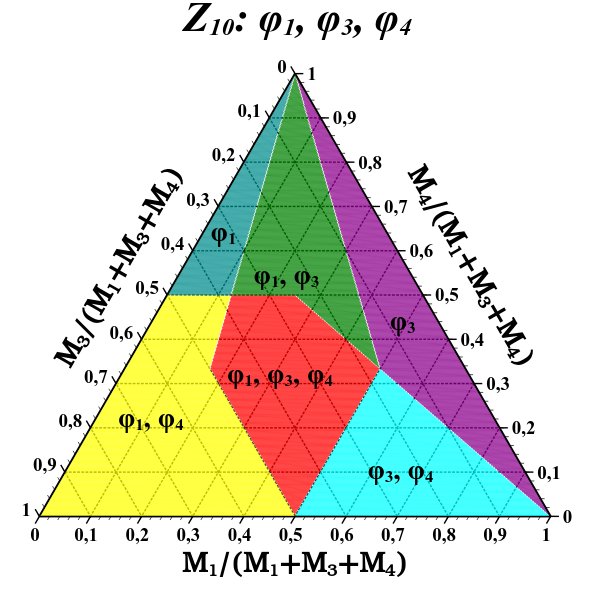}\caption{Stability regions for a $Z_9$ (left panel) and  $Z_{10}$ (right panel) symmetry with fields  $\phi_1,\phi_3,\phi_4$.}
\label{fig:trianglep134}
\end{figure}
\item $(\phi_2,\phi_3,\phi_4)$:  The interaction terms are
\begin{align}
 \mathcal{L}_3\supset&\,\, \phi _3^3,\phi _2 \phi _3 \phi _4,\phi _2^2 \phi _{\text{s4}}.\\
  \mathcal{L}_4\supset&\,\, \phi _2^3 \phi _3,\phi _3 \phi _4^2 \phi _{\text{s2}},\phi _2^2 \phi
   _{\text{s2}}^2,\phi _4^3 \phi _{\text{s3}},\phi _2 \phi _3 \phi _{\text{s2}} \phi
   _{\text{s3}},\phi _3^2 \phi _{\text{s3}}^2,\phi
   _2 \phi _4 \phi _{\text{s3}}^2,\nonumber\\
   &\phi _2
   \phi _4 \phi _{\text{s2}} \phi _{\text{s4}},\phi _3 \phi _4 \phi _{\text{s3}} \phi
   _{\text{s4}},\phi _4^2 \phi _{\text{s4}}^2.\\
   \mathcal{L}_5\supset&\,\, \phi _2 \phi _4^4,\phi _2^2 \phi _4^2 \phi _{\text{s3}},\phi _2 \phi _3^2 \phi _{\text{s4}}^2.
 \end{align}
 
 These interactions give rise to two- and three-body decays for every particle. The three-particle stability region, described by $M_4<2 M_2,\, M_2<M_3+M_4,\, M_3<M_2+M_4,\, M_4<M_2+M_3,\, M_3<3 M_4$, 
% $\left(\frac{M_2}{4}<M_4\leq \frac{M_2}{2}\land M_2-M_4<M_3<3M_4\right)\,\,\lor$\\ $\left(\frac{M_2}{2}<M_4\leq M_2\land M_2-M_4<M_3<M_2+M_4\right)\,\,\lor$ $\left(M_2<M_4<2 M_2\land M_4-M_2<M_3<M_2+M_4\right)$, 
 is shown in the  left panel of Fig.~\ref{fig:trianglep234}  (red region). That figure also displays the other 5 possibilities  regarding stability.
 \end{itemize}

\begin{figure}
\centering
\includegraphics[scale=0.5]{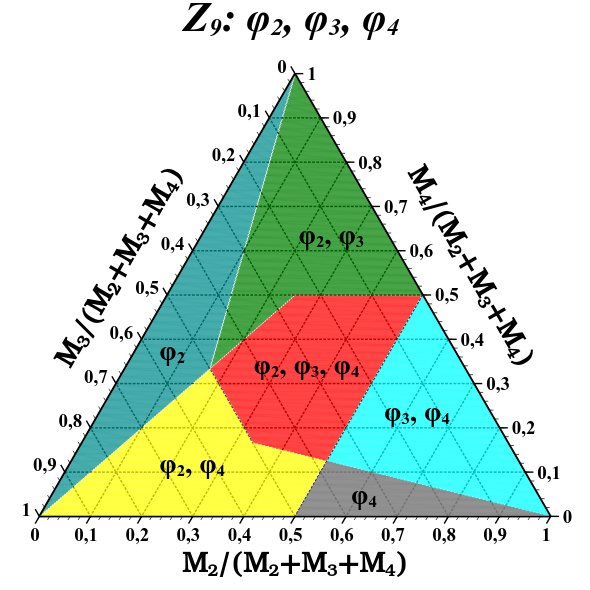}\includegraphics[scale=0.5]{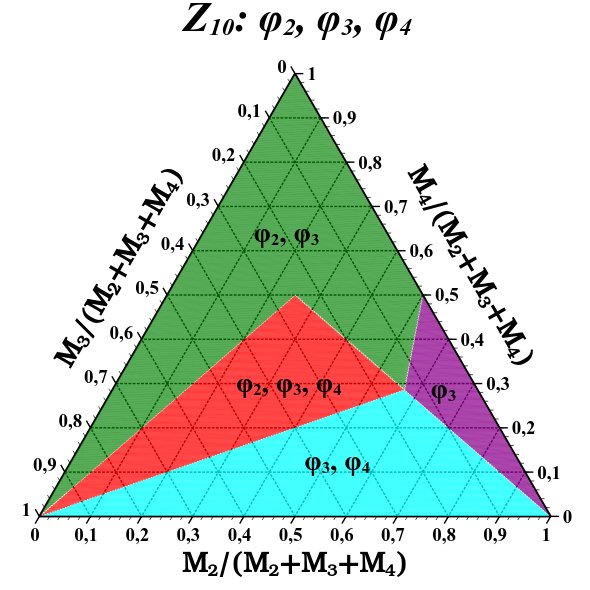}\caption{Stability regions for a $Z_9$ (left panel) and  $Z_{10}$ (right panel) symmetry with fields  $\phi_2,\phi_3,\phi_4$.}
\label{fig:trianglep234}
\end{figure}
\subsubsection{Four fields}
\begin{figure}
\centering
\includegraphics[scale=0.5]{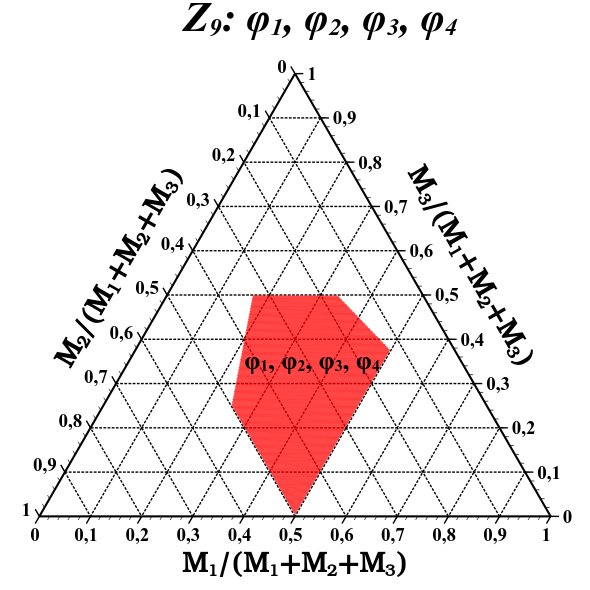}\includegraphics[scale=0.5]{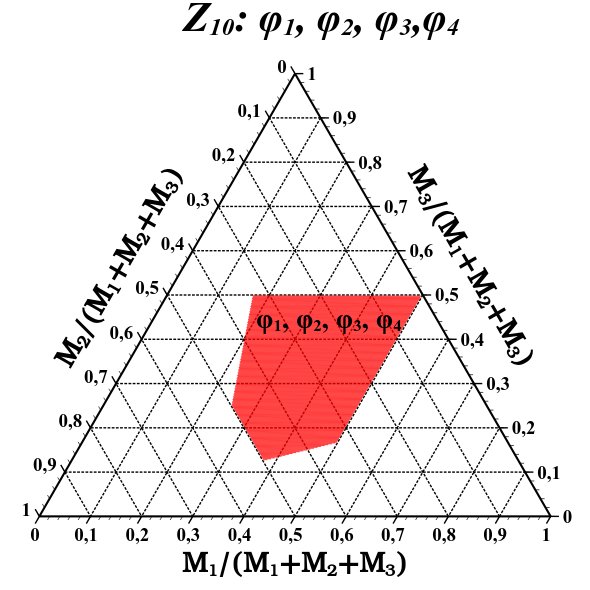}\caption{The region where the four fields $\phi_1,\phi_2,\phi_3,\phi_4$ can be stable for a $Z_9$ (left panel) and  a $Z_{10}$ (right panel) symmetry. Inside the colored region it is possible to find a value of $M_4$ such that all four fields are stable.}
\label{fig:trianglep1234}
\end{figure}

The unique scenario with four fields charged under a $Z_9$ symmetry has the following interaction terms:
\begin{align}
 \mathcal{L}_3\supset&\,\, \phi _3^3,\phi _2 \phi _3 \phi _4,\phi _1 \phi _4^2,\phi _1^2 \phi _{\text{s2}},\phi _4 \phi _{\text{s2}}^2,\phi _1 \phi _2 \phi _{\text{s3}},\phi _1 \phi _3 \phi _{\text{s4}}.\\
  \mathcal{L}_4\supset&\,\, \phi _2^3 \phi _3,\phi _1 \phi _2 \phi _3^2,\phi _1 \phi _2^2 \phi _4,\phi _1^2
   \phi _3 \phi _4,\phi _3^2 \phi _4 \phi _{\text{s1}},\phi _2 \phi _4^2 \phi
   _{\text{s1}},\phi _1^2 \phi _{\text{s1}}^2,\phi _3 \phi _{\text{s1}}^3,\phi _3 \phi
   _4^2 \phi _{\text{s2}},\phi _1 \phi _2 \phi _{\text{s1}} \phi _{\text{s2}},\nonumber\\
   &\phi _4
   \phi _{\text{s1}}^2 \phi _{\text{s2}},\phi _2^2 \phi _{\text{s2}}^2,\phi _1 \phi _3
   \phi _{\text{s2}}^2,\phi _4^3 \phi _{\text{s3}},\phi _1 \phi _3 \phi _{\text{s1}} \phi
   _{\text{s3}},\phi _2 \phi _3 \phi _{\text{s2}} \phi _{\text{s3}},\phi _1 \phi _4 \phi
   _{\text{s2}} \phi _{\text{s3}},\nonumber\\
   &\phi _3^2 \phi
   _{\text{s3}}^2,\phi _2 \phi _4 \phi _{\text{s3}}^2,\phi _1 \phi _4 \phi _{\text{s1}} \phi
   _{\text{s4}},\phi _2 \phi _4 \phi
   _{\text{s2}} \phi _{\text{s4}},\phi _3 \phi _4 \phi _{\text{s3}} \phi _{\text{s4}},\phi _4^2 \phi _{\text{s4}}^2.\\
   \mathcal{L}_5\supset&\,\, \phi _1 \phi _2^4,\phi _1^2 \phi _2^2 \phi _3,\phi _1^3 \phi _2 \phi _4,\phi _2
   \phi _4^4,\phi _2^2 \phi _3^2 \phi _{\text{s1}},\phi _2^3 \phi _4 \phi
   _{\text{s1}},\phi _3 \phi _4^2 \phi _{\text{s1}}^2,\phi _4 \phi _{\text{s1}}^4,\phi
   _1 \phi _3^2 \phi _4 \phi _{\text{s2}},\phi _4^3 \phi _{\text{s1}} \phi
   _{\text{s2}},\nonumber\\
   &\phi _2^2 \phi _4^2 \phi
   _{\text{s3}},\phi _1^2 \phi
   _4 \phi _{\text{s3}}^2,\phi _4^2 \phi _{\text{s2}} \phi _{\text{s3}}^2.
 \end{align}
The conditions that ensure the stability of the four particles at the renormalizable level are 
$M_2<2 M_1,\, M_4<2 M_2,\, M_1<2 M_4,\, M_1<M_2+M_3,\, M_2<M_1+M_3,\,
   M_3<M_1+M_2,\, M_1<M_3+M_4,\, M_3<M_1+M_4,\, M_4<M_1+M_3,\, M_2<M_3+M_4,\,
   M_3<M_2+M_4,\, M_4<M_2+M_3$. Since we now deal with four particles, it becomes more difficult to illustrate graphically this region of parameter space. In the left panel of Fig.~\ref{fig:trianglep1234} we display, via a ternary plot, the region where a four-component dark matter scenario \emph{could} be attained. Within the red region, there exists values of $M_4$ (not shown) such that all four particles are stable. Let us stress that this does not mean that all four particles will be stable for an arbitrary value of $M_4$.
%$M_2<2 M_1\land M_4<2 M_2\land M_1<2 M_4\land M_1<M_2+M_3\land M_2<M_1+M_3\land M_3<M_1+M_2\land M_1<M_3+M_4\land M_3<M_1+M_4\land M_4<M_1+M_3\land M_2<M_3+M_4\land M_3<M_2+M_4\land M_4<M_2+M_3$. 
   This is the minimal setup that allows to realize a four-component dark matter scenario with  complex scalar fields.

\subsection{$Z_{10}$}
Under this symmetry scenarios with up to  four different  fields arise. For concreteness, we only discuss the scenarios with three and four fields. 

\subsubsection{Three fields}

\begin{itemize}
    \item $(\phi_1,\phi_2,\phi_3)$: The interaction terms are
\begin{align}
 \mathcal{L}_3\supset&\,\, \phi _1^2 \phi _{\text{s2}},\phi _1 \phi _2 \phi _{\text{s3}}.\\
  \mathcal{L}_4\supset&\,\, \phi _2^2 \phi _3^2,\phi _1 \phi _3^3,\phi _1^2 \phi _{\text{s1}}^2,\phi _3 \phi
   _{\text{s1}}^3,\phi _1 \phi _2 \phi _{\text{s1}} \phi _{\text{s2}},\phi _2^2 \phi
   _{\text{s2}}^2,\phi _1 \phi _3 \phi _{\text{s2}}^2,\nonumber\\
   &\phi _1 \phi _3 \phi _{\text{s1}} \phi
   _{\text{s3}},\phi _2 \phi _3 \phi _{\text{s2}} \phi _{\text{s3}},\phi _3^2 \phi
   _{\text{s3}}^2.\\
   \mathcal{L}_5\supset&\,\, \phi _1 \phi _2^3 \phi _3,\phi _1^2 \phi _2 \phi _3^2,\phi _2 \phi _3^3 \phi
   _{\text{s1}},\phi _3^4 \phi _{\text{s2}}.
 \end{align}
In this scenario $\phi_2$ has only two-body decays whereas the other two fields have two- and three-body decays. The full stability region, described by the condition  
$M_2<2 M_1,\, M_1<M_2+M_3,\, M_2<M_1+M_3,\, M_3<M_1+M_2,\, M_1<3 M_3$,
%$\left(\frac{M_1}{3}<M_3\leq M_1\land\right.$ $\left.M_1-M_3<M_2<M_1+M_3\right)\,\,\lor$ $\left(M_1<M_3<3 M_1\land M_3-M_1<M_2<2 M_1\right)$,
is shown in the bottom right panel of Fig.~\ref{fig:trianglep123}, which also displays the other possibilities.
  
\item $(\phi_1,\phi_2,\phi_4)$:  The interaction terms are
\begin{align}
 \mathcal{L}_3\supset&\,\, \phi _2 \phi _4^2,\phi _1^2 \phi _{\text{s2}},\phi _2^2 \phi _{\text{s4}}.\\
  \mathcal{L}_4\supset&\,\, \phi _2^3 \phi _4,\phi _1^2 \phi _4^2,\phi _1^2 \phi _{\text{s1}}^2,\phi _4^3
   \phi _{\text{s2}},\phi _1 \phi _2 \phi _{\text{s1}} \phi _{\text{s2}},\phi _4 \phi
   _{\text{s1}}^2 \phi _{\text{s2}},\phi _2^2 \phi _{\text{s2}}^2,\phi _1 \phi _4 \phi _{\text{s1}} \phi _{\text{s4}},\nonumber\\
   &\phi _2 \phi _4 \phi
   _{\text{s2}} \phi _{\text{s4}},\phi _4^2 \phi
   _{\text{s4}}^2.\\
   \mathcal{L}_5\supset&\,\, \phi _1^2 \phi _2^2 \phi _4,\phi _1^4 \phi
   _{\text{s4}}.
 \end{align}
Consequently, $\phi_1$ is always stable even at the non-renormalizable level ($d=5$). 
The full stability region, described by the condition $M_2<2 M_1,\, M_4<2 M_2,\, M_2<2 M_4$, 
%$M_2<2 M_1\land \frac{M_2}{2}<M_4<2 M_2$, 
is illustrated in the right panel of Fig.~\ref{fig:trianglep124}, which also shows the remaining 3 cases. 

\item $(\phi_1,\phi_3,\phi_4)$:  The interaction terms are
\begin{align}
 \mathcal{L}_3\supset&\,\, \phi _3^2 \phi _4,\phi _1 \phi _3
   \phi _{\text{s4}}.\\
  \mathcal{L}_4\supset&\,\, \phi _1 \phi _3^3,\phi _1^2 \phi _4^2,\phi _3 \phi _4^2 \phi _{\text{s1}},\phi
   _1^2 \phi _{\text{s1}}^2,\phi _1^3 \phi _{\text{s3}},\phi
   _1 \phi _3 \phi _{\text{s1}} \phi _{\text{s3}},\phi _3^2 \phi _{\text{s3}}^2,\nonumber\\
   &\phi _1 \phi _4 \phi _{\text{s1}} \phi
   _{\text{s4}},\phi _3 \phi _4 \phi _{\text{s3}} \phi _{\text{s4}},\phi _4^2 \phi
   _{\text{s4}}^2.\\
   \mathcal{L}_5\supset&\,\, \phi _1^3 \phi _3 \phi _4,\phi _4 \phi _{\text{s1}}^4,\phi _1 \phi _4^3 \phi
   _{\text{s3}},\phi _1^2 \phi _4 \phi _{\text{s3}}^2.
 \end{align}
It follows that the condition $M_4<2 M_3,\, M_1<M_3+M_4,\, M_3<M_1+M_4,\, M_4<M_1+M_3,\, M_3<3 M_1$,
%$\left(\frac{M_1}{3}<M_3\leq M_1\land M_1-M_3<M_4<2 M_3\right)\,\,\lor$\\ $\left(M_1<M_3<3 M_1\land M_3-M_1<M_4<M_1+M_3\right)$ 
avoids all the two- and three- body decays and leads to the stability (red) region shown in the right panel of Fig.~\ref{fig:trianglep134}.  

\item $(\phi_2,\phi_3,\phi_4)$:  The interaction terms are
\begin{align}
 \mathcal{L}_3\supset&\,\, \phi _3^2 \phi _4,\phi _2 \phi _4^2,\phi _2^2 \phi
   _{\text{s4}}.\\
  \mathcal{L}_4\supset&\,\, \phi _2^2 \phi _3^2,\phi _2^3 \phi _4,\phi _4^3 \phi _{\text{s2}},\phi _2^2 \phi
   _{\text{s2}}^2,\phi _2 \phi _3 \phi _{\text{s2}} \phi _{\text{s3}},\phi _3^2 \phi
   _{\text{s3}}^2,\phi _2 \phi _4 \phi _{\text{s3}}^2,\nonumber\\
   &\phi _2 \phi _4 \phi
   _{\text{s2}} \phi _{\text{s4}},\phi _3 \phi _4
   \phi _{\text{s3}} \phi _{\text{s4}},\phi _4^2 \phi _{\text{s4}}^2.\\
   \mathcal{L}_5\supset&\,\, \phi
   _2 \phi _{\text{s3}}^4,\phi _2 \phi _3^2 \phi _{\text{s4}}^2.
 \end{align}
 Notice that $\phi_3$ is always stable, even at the non-renormalizable level. The full stability region is ensured by the condition $M_4<2 M_2,\, M_4<2 M_3,\, M_2<2 M_4$ 
 %$\frac{M_4}{2}<M_2<2 M_4\land M_3>\frac{M_4}{2}$,
 and corresponds to the red region shown in the right panel of Fig.~\ref{fig:trianglep234}.  The other three stability regions are also displayed in that same figure.

\end{itemize}
\subsubsection{Four fields}
The scenario with the four fields $(\phi_1,\phi_2,\phi_3,\phi_4)$ features the following interaction terms:
\begin{align}
 \mathcal{L}_3\supset&\,\, \phi _3^2 \phi _4,\phi _2 \phi _4^2,\phi _1^2 \phi
   _{\text{s2}},\phi _1 \phi _2 \phi _{\text{s3}},\phi _2^2 \phi _{\text{s4}},\phi _1 \phi _3 \phi _{\text{s4}}.\\
  \mathcal{L}_4\supset&\,\, \phi _2^2 \phi _3^2,\phi _1 \phi _3^3,\phi _2^3 \phi _4,\phi _1 \phi _2 \phi _3
   \phi _4,\phi _1^2 \phi _4^2,\phi _3 \phi _4^2 \phi _{\text{s1}},\phi _1^2 \phi
   _{\text{s1}}^2,\phi _3 \phi _{\text{s1}}^3,\phi _4^3 \phi _{\text{s2}},\phi _1 \phi
   _2 \phi _{\text{s1}} \phi _{\text{s2}},\phi _4 \phi _{\text{s1}}^2 \phi
   _{\text{s2}},\nonumber\\
   &\phi _2^2 \phi _{\text{s2}}^2,\phi _1 \phi _3 \phi _{\text{s2}}^2,\phi _1 \phi _3
   \phi _{\text{s1}} \phi _{\text{s3}},\phi _2 \phi _3 \phi _{\text{s2}} \phi
   _{\text{s3}},\phi _1 \phi _4 \phi _{\text{s2}} \phi _{\text{s3}},\phi _3^2 \phi
   _{\text{s3}}^2,\phi _2 \phi _4 \phi _{\text{s3}}^2,\nonumber\\
   &\phi _1 \phi _4 \phi
   _{\text{s1}} \phi _{\text{s4}},\phi _2
   \phi _4 \phi _{\text{s2}} \phi _{\text{s4}},\phi _3 \phi _4 \phi _{\text{s3}} \phi _{\text{s4}}.\\
   \mathcal{L}_5\supset&\,\, \phi _1 \phi _2^3 \phi _3,\phi _1^2 \phi _2 \phi _3^2,\phi _1^2 \phi _2^2 \phi
   _4,\phi _1^3 \phi _3 \phi _4,\phi _2 \phi _3^3 \phi _{\text{s1}},\phi _2^2 \phi _3
   \phi _4 \phi _{\text{s1}},\phi _4 \phi _{\text{s1}}^4,\phi _3^4 \phi
   _{\text{s2}},\phi _1 \phi _3 \phi _4^2 \phi _{\text{s2}},\phi _1 \phi _4^3 \phi
   _{\text{s3}},\nonumber\\
   &\phi _1^2 \phi
   _4 \phi _{\text{s3}}^2,\phi _4^2 \phi _{\text{s2}} \phi _{\text{s3}}^2.
 \end{align}

It follows that the full stability region is given by the condition $M_2<2 M_1\land M_4<2 M_2\land M_4<2 M_3\land M_2<2 M_4\land M_1<M_2+M_3\land
   M_2<M_1+M_3\land M_3<M_1+M_2\land M_1<M_3+M_4\land M_3<M_1+M_4\land M_4<M_1+M_3$ --see the right panel of  Fig. \ref{fig:trianglep1234}. 
   
   This is the last case we are going to examine for complex dark matter. It is clear, though, that the discussion can be extended to even higher $N$. Notice that for the dark matter to consists of $k$ complex particles stabilized with a single $Z_N$ symmetry, $N$ must at least be  $2k+1$.

\section{Stability analysis: Complex and Real Dark Matter}
\label{sec:real}

When  $N$ is even and the field $\phi_{N/2}$ is  present a novel situation arises that leads to a real dark matter particle. In fact, the quadratic term $\phi_{N/2}^2+\text{h.c.}$  is also invariant under the  $Z_N$ and splits the complex field $\phi_{N/2}$ into two \emph{real} fields with different masses. These two mass eigenstates are thus linear combinations of $\phi_{N/2}$ and $\phi_{N/2}^\dagger$, and do not have a definite charge under $Z_N$. Moreover, the heavier of them would necessarily decay, via the term $(\phi_{N/2}^2+\text{h.c.})H^\dagger H$, into the lighter one plus SM particles, so that only the lighter one can be stable. Whether it is really stable or not will depend on the allowed interactions with the other scalar fields charged under $Z_N$ and on the relations among their masses.

Let us denote the lighter mass eigenstate, which is a real field, by $\phi'_{N/2}$ and its mass by $M'_{N/2}$. Then, the stability conditions can be read off directly from the Lagrangian  in a way completely analogous to that for complex fields --see previous section-- but will involve restrictions on $M'_{N/2}$.  In this section, we consider only the cases where $N$ is even and the field $\phi_{N/2}$ is present. They lead to multi-component dark matter scenarios in which one (and only one) of the dark matter particles is a real scalar field ($\phi'_{N/2}$) while the rest are complex scalar fields. Notice that it was recently pointed out \cite{Queiroz:2016sxf, Kavanagh:2017hcl} that it is feasible to experimentally distinguish between a real and a complex dark matter particle.  Next we analyze the possible scenarios for different $Z_N$.

\subsection{$Z_4$}
We can only have two different fields charged under the symmetry, $\phi_1$ and $\phi_2$.
The interaction terms are
\begin{align}
 \mathcal{L}_3\supset&\,\, \phi _1^2 \phi _2,\phi _1^2 \phi _{\text{s2}}.\\
  \mathcal{L}_4\supset&\,\,  \phi _1^4,\phi _2^4,\phi _1 \phi _2^2 \phi _{\text{s1}},\phi _1^2 \phi _{\text{s1}}^2,\phi _2^3 \phi
   _{\text{s2}},\phi _1 \phi _2 \phi _{\text{s1}} \phi _{\text{s2}},\phi _2^2 \phi _{\text{s2}}^2.%\\
%   \mathcal{L}_5\supset&\,\, 0.
 \end{align}
In this case the Lagrangian term $\phi_2^\dagger\phi_1^2$ is allowed, which entails that $\phi_1$ is automatically stable while $\phi'_2$ (the lighter mass eigenstate) will be stable for $M'_2<2M_1$. There are no $d=5$ operators inducing $\phi_1$ or $\phi'_2$ decays. Hence, for $M'_2<2M_1$ the dark matter would consist of two particles: one complex scalar ($\phi_1$) and one real scalar ($\phi'_2$). Models similar to this one were considered in \cite{Batell:2010bp,Belanger:2014bga}.

\subsection{$Z_6$}
We can either have two or three different fields charged under the symmetry: $\phi_1$, $\phi_2$ and $\phi_3$. 
\subsubsection{Two fields}
\begin{itemize}
 \item  $(\phi_1,\phi_3)$: The interaction terms are
\begin{align}
% \mathcal{L}_3=&\,\, 0.\\
  \mathcal{L}_4\supset&\,\, \phi _1^3 \phi _3,\phi _3^4,\phi _1 \phi _3^2 \phi _{\text{s1}},\phi _1^2 \phi _{\text{s1}}^2,\phi _1^3 \phi _{\text{s3}},\phi _3^3 \phi _{\text{s3}},\phi _1 \phi _3 \phi _{\text{s1}} \phi
   _{\text{s3}},\phi _3^2 \phi _{\text{s3}}^2.%\\
%   \mathcal{L}_5\supset&\,\, 0.
 \end{align}
   Thus, $\phi_1$ is automatically stable while $\phi'_3$ will be stable for $M'_3<3M_1$. 
\item $(\phi_2,\phi_3)$: The interaction terms are
\begin{align}
 \mathcal{L}_3\supset&\,\, \phi _2^3.\\
  \mathcal{L}_4\supset&\,\, \phi _3^4,\phi _2 \phi _3^2 \phi _{\text{s2}},\phi _2^2 \phi _{\text{s2}}^2,\phi _3^3 \phi _{\text{s3}},\phi _2 \phi _3
   \phi _{\text{s2}} \phi _{\text{s3}},\phi _3^2 \phi _{\text{s3}}^2.%\\
%   \mathcal{L}_5\supset&\,\, 0.
 \end{align}
In this case there are neither cubic nor quartic terms involving one single field. Thus, both fields, $\phi_2$ and $\phi'_3$, are stable independently of their masses. We refer to this situation as \emph{unconditional stability}. A $Z_6$ symmetry with fields $\phi_{2,3}$ is the simplest scenario in which unconditional stability arises. Moreover, as explained in the appendix \ref{sec:unco}, in this case unconditional stability is not limited to the renormalizable Lagrangian but is maintained for operators of arbitrary dimension.  A related model was mentioned in \cite{Batell:2010bp}.
\end{itemize}
\subsubsection{Three fields}
\begin{figure}
\centering
\includegraphics[scale=0.5]{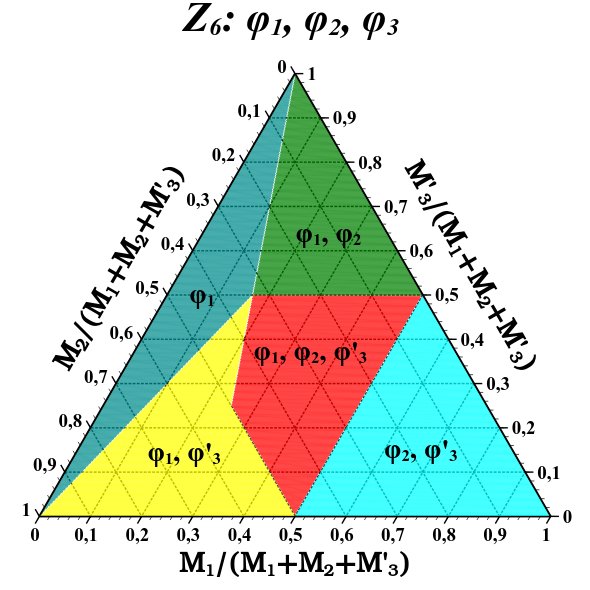}\includegraphics[scale=0.5]{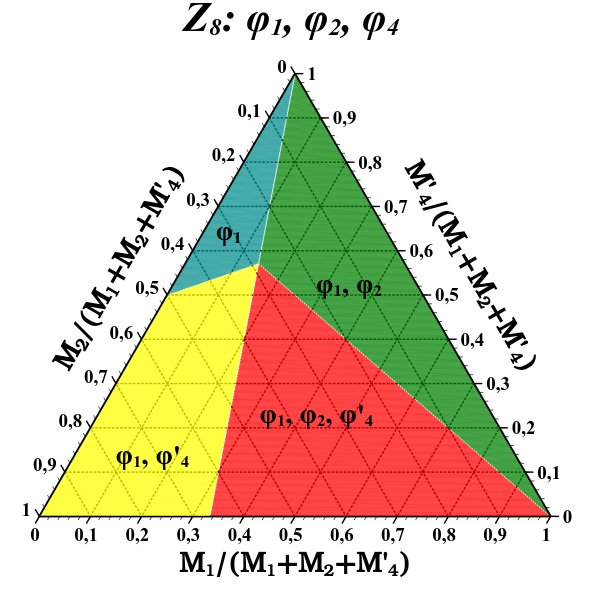}\\
\vspace{0.5cm}
\includegraphics[scale=0.5]{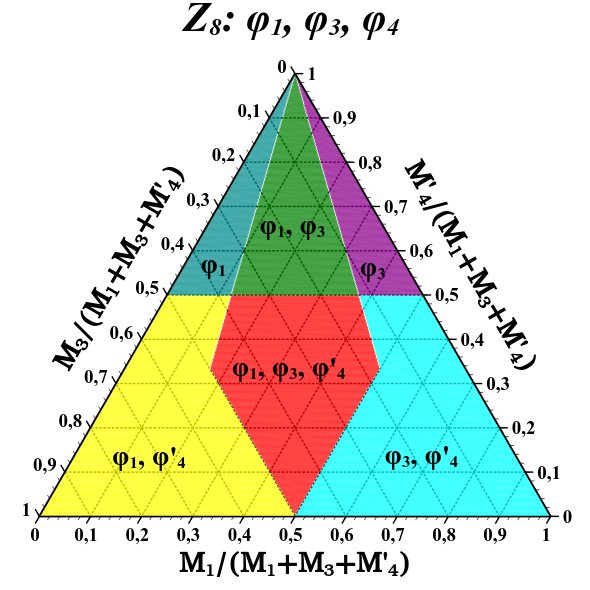}\includegraphics[scale=0.5]{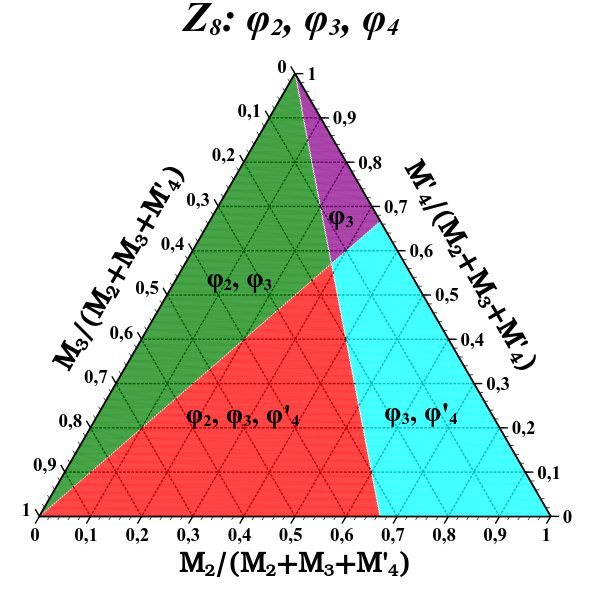}\caption{Stability regions for a $Z_6$ with fields $\phi_1,\phi_2,\phi_3$ (top left panel), and for a $Z_8$ with different sets of fields: $\phi_1,\phi_2,\phi_4$ (top right panel), $\phi_1,\phi_3,\phi_4$ (bottom left panel) and $\phi_2,\phi_3,\phi_4$ (bottom right panel). For the $Z_6$ ($Z_8$) symmetry the field $\phi_3$ ($\phi_4$) denotes the real part of the corresponding field.  }
%\label{fig:trianglez7}
\label{fig:trianglez6z8}
\end{figure}

The only possibility is $(\phi_1,\phi_2,\phi_3)$ with interaction terms given by
\begin{align}
 \mathcal{L}_3\supset&\,\, \phi _2^3,\phi _1 \phi _2 \phi _3,\phi _1^2 \phi _{\text{s2}},\phi _1 \phi _2 \phi _{\text{s3}}.\\
  \mathcal{L}_4\supset&\,\, \phi _1^2 \phi _2^2,\phi _1^3 \phi _3,\phi _3^4,\phi _2^2 \phi _3 \phi _{\text{s1}},\phi _1 \phi _3^2 \phi
   _{\text{s1}},\phi _1^2 \phi _{\text{s1}}^2,\phi _3 \phi _{\text{s1}}^3,\phi _2 \phi _3^2 \phi _{\text{s2}},\phi _1 \phi _2
   \phi _{\text{s1}} \phi _{\text{s2}},\phi _2^2 \phi _{\text{s2}}^2,\phi _1 \phi _3 \phi _{\text{s2}}^2,\nonumber\\
   &\phi _3^3 \phi _{\text{s3}},\phi _1 \phi _3 \phi _{\text{s1}} \phi _{\text{s3}},\phi _2 \phi _3 \phi
   _{\text{s2}} \phi _{\text{s3}},\phi _3^2 \phi _{\text{s3}}^2.\\
   \mathcal{L}_5\supset&\,\, \phi _1^4 \phi _2,\phi _1 \phi _2 \phi _3^3,\phi _2 \phi _3^2 \phi
   _{\text{s1}}^2,\phi _1^2 \phi _3^2 \phi _{\text{s2}},\phi _3^3 \phi _{\text{s1}} \phi
   _{\text{s2}}.
 \end{align}
Hence, the three fields will be stable for $M_2<2 M_1,\, M_1<M_2+M'_3,\, M_2<M_1+M'_3,\, M'_3<M_1+M_2$.  
%$\left(M_2\leq M_1\land M_1-M_2<M'_3<M_1+M_2\right)\,\,\lor$\\ $\left(M_1<M_2<2 M_1\land M_2-M_1<M'_3<M_1+M_2\right)$. 
In that case, the dark matter would consists of three particles: two complex scalar fields ($\phi_{1,2}$) and one real scalar field ($\phi'_3$) \cite{Batell:2010bp}. The five possible stability regions for this case are displayed in the top left panel of Figure \ref{fig:trianglez6z8}.

\subsection{$Z_8$}
We can either have two, three or four different fields charged under $Z_8$. %the symmetry: $\phi_1$, $\phi_2$,  $\phi_3$ and $\phi_4$.
\subsubsection{Two fields}
\begin{itemize}
\item $(\phi_1,\phi_4)$: The interaction terms are
\begin{align}
% \mathcal{L}_3\supset&\,\, 0 .\\
  \mathcal{L}_4\supset&\,\, \phi _4^4,\phi _1 \phi _4^2 \phi _{\text{s1}},\phi _1^2 \phi _{\text{s1}}^2,\phi _4^3 \phi _{\text{s4}},\phi _1 \phi _4
   \phi _{\text{s1}} \phi _{\text{s4}},\phi _4^2 \phi _{\text{s4}}^2.\\
  \mathcal{L}_5\supset&\,\, \phi _1^4 \phi _4,\phi _1^4 \phi _{\text{s4}}.
 \end{align}
Notice that at the renormalizable level both particles, $\phi_1$ and $\phi'_4$, are automatically stable. This is another example of unconditional stability but limited to the renormalizable Lagrangian. In fact, the $d=5$ interactions, if present, would induce the decay $\phi'_4\to 4\phi_1$ for $M_4'>4M_1$. 

    \item $(\phi_2,\phi_4)$:   The interaction terms are
\begin{align}
 \mathcal{L}_3\supset&\,\, \phi _2^2 \phi _4,\phi _2^2 \phi _{\text{s4}}.\\
  \mathcal{L}_4\supset&\,\, \phi _2^4,\phi _4^4,\phi _2 \phi _4^2 \phi _{\text{s2}},\phi _2^2 \phi _{\text{s2}}^2,\phi _4^3 \phi
   _{\text{s4}},\phi _2 \phi _4 \phi _{\text{s2}} \phi _{\text{s4}},\phi _4^2 \phi _{\text{s4}}^2.%\\
%  \mathcal{L}_5\supset&\,\, 0.
 \end{align}
In this case $\phi_2$ is automatically stable while $\phi'_4$ will be stable for $M'_4<2M_2$. There are no $d=5$ operators inducing decays. 
    \item $(\phi_3,\phi_4)$:   The interaction terms are
\begin{align}
% \mathcal{L}_3\supset&\,\, 0 .\\
  \mathcal{L}_4\supset&\,\, \phi _4^4,\phi _3 \phi _4^2 \phi _{\text{s3}},\phi _3^2 \phi _{\text{s3}}^2,\phi _4^3 \phi _{\text{s4}},\phi _3 \phi _4
   \phi _{\text{s3}} \phi _{\text{s4}},\phi _4^2 \phi _{\text{s4}}^2.\\
  \mathcal{L}_5\supset&\,\, \phi _3^4 \phi _4,\phi _3^4 \phi _{\text{s4}}.
 \end{align}
Here we have another example of unconditional stability, for $\phi_3$ and $\phi'_4$, at the renormalizable level. The $d=5$ operators, if present, would induce the decay $\phi'_4\to 4\phi_3$ for $M_4'>4M_3$.

\end{itemize}
\subsubsection{Three fields}
\begin{itemize}
     \item  $(\phi_1,\phi_2,\phi_4)$: The interaction terms are
\begin{align}
 \mathcal{L}_3\supset&\,\, \phi _2^2 \phi _4,\phi _1^2 \phi _{\text{s2}},\phi _2^2 \phi
   _{\text{s4}} .\\
  \mathcal{L}_4\supset&\,\, \phi _2^4,\phi _1^2 \phi _2 \phi _4,\phi _4^4,\phi _1 \phi _4^2 \phi _{\text{s1}},\phi _1^2 \phi _{\text{s1}}^2,\phi _2
   \phi _4^2 \phi _{\text{s2}},\phi _1 \phi _2 \phi _{\text{s1}} \phi _{\text{s2}},\phi _2^2 \phi _{\text{s2}}^2,\phi _1^2 \phi _2 \phi _{\text{s4}},\nonumber\\
   &\phi _4^3 \phi
   _{\text{s4}},\phi _1 \phi _4 \phi _{\text{s1}} \phi _{\text{s4}},
   \phi _2 \phi _4 \phi _{\text{s2}} \phi _{\text{s4}},\phi _4^2 \phi _{\text{s4}}^2.\\
  \mathcal{L}_5\supset&\,\, \phi _1^4 \phi _4,\phi _2 \phi _4^2 \phi _{\text{s1}}^2,\phi _4 \phi
   _{\text{s1}}^4,\phi _1^2 \phi _4^2 \phi _{\text{s2}}.
 \end{align}
The stability condition for all three particles ($\phi_{1},\phi_2,\phi'_4$) is $M_2<2 M_1,\, M'_4<2 M_2$, 
which leads to the red region displayed in the top right panel of Figure \ref{fig:trianglez6z8}. For this set of charges (fields), it is not possible to write an invariant term linear in $\phi_1$. Consequently, $\phi_1$ is always stable and only four stability regions are found. 
\item $(\phi_1,\phi_3,\phi_4)$: The interaction terms are
\begin{align}
 \mathcal{L}_3\supset&\,\, \phi _1 \phi _3 \phi _4,\phi _1 \phi _3 \phi _{\text{s4}} .\\
  \mathcal{L}_4\supset&\,\, \phi _1^2 \phi _3^2,\phi _4^4,\phi _3^3 \phi _{\text{s1}},\phi _1 \phi _4^2 \phi _{\text{s1}},\phi _1^2 \phi
   _{\text{s1}}^2,\phi _1^3 \phi _{\text{s3}},\phi _3 \phi _4^2 \phi _{\text{s3}},\phi _1 \phi _3
   \phi _{\text{s1}} \phi _{\text{s3}},\phi _3^2 \phi _{\text{s3}}^2,\nonumber\\
   & \phi _4^3 \phi _{\text{s4}},\phi _1 \phi _4 \phi _{\text{s1}} \phi _{\text{s4}},\phi _3 \phi _4 \phi
   _{\text{s3}} \phi _{\text{s4}},\phi _4^2 \phi _{\text{s4}}^2.\\
  \mathcal{L}_5\supset&\,\, \phi _1^4 \phi _4,\phi _3^4 \phi _4,\phi _1 \phi _3 \phi _4^3,\phi _3^2 \phi _4
   \phi _{\text{s1}}^2,\phi _4 \phi _{\text{s1}}^4,\phi _4^3 \phi _{\text{s1}} \phi
   _{\text{s3}},\phi _1^2 \phi _4 \phi _{\text{s3}}^2,\phi _4 \phi _{\text{s3}}^4.
 \end{align}
The three particles are stable as long as $M_1<M_3+M'_4,\, M_3<M_1+M'_4,\, M'_4<M_1+M_3,\, M_1<3 M_3,\, M_3<3 M_1$.
%$\left(\frac{M_1}{3}<M_3\leq M_1\land M_1-M_3<M'_4<M_1+M_3\right)\,\,\lor$ $\left(M_1<M_3<3 M_1\land M_3-M_1<M'_4<M_1+M_3\right)$. 
The stability regions for this case are displayed in the bottom left panel of Figure \ref{fig:trianglez6z8}. 
%  It follows that $M_i<M_j+M_k$, $M_1<3M_3$, $M_3<3M_1$,
  \item  $(\phi_2,\phi_3,\phi_4)$: The interaction terms are
\begin{align}
 \mathcal{L}_3\supset&\,\, \phi _2 \phi _3^2,\phi _2^2 \phi _4,\phi _4 \phi _{\text{s2}}^2.\\
  \mathcal{L}_4\supset&\,\, \phi _2^4,\phi _4^4,\phi _3^2 \phi _4 \phi _{\text{s2}},\phi _2 \phi _4^2 \phi _{\text{s2}},\phi _2^2 \phi
   _{\text{s2}}^2,\phi _3 \phi _4^2 \phi _{\text{s3}},\phi _2 \phi _3 \phi _{\text{s2}} \phi
   _{\text{s3}},\phi _3^2 \phi _{\text{s3}}^2,\phi _2 \phi _4 \phi _{\text{s3}}^2,\phi _4^3 \phi _{\text{s4}},\nonumber\\
   &\phi _2 \phi _4 \phi _{\text{s2}} \phi _{\text{s4}},\phi _3 \phi _4 \phi _{\text{s3}} \phi
   _{\text{s4}},\phi _4^2 \phi _{\text{s4}}^2.\\
  \mathcal{L}_5\supset&\,\, \phi _3^4 \phi _4,\phi _2 \phi _3^2 \phi _4^2,\phi _4^2 \phi _{\text{s2}} \phi
   _{\text{s3}}^2,\phi _4 \phi _{\text{s3}}^4 .
 \end{align}
 The stability regions for this case are displayed in the bottom right panel of Figure \ref{fig:trianglez6z8}. Notice that $\phi_3$ is always stable. The red region, in which $\phi_{2,3}$ and $\phi'_4$ are simultaneously stable, is obtained by the condition $M_2<2 M_3,\, M'_4<2 M_2$. %$M_3>\frac{M_2}{2}\land M'_4<2 M_2$. 

\end{itemize}
\subsubsection{Four fields}
\begin{figure}
\centering
\includegraphics[scale=0.5]{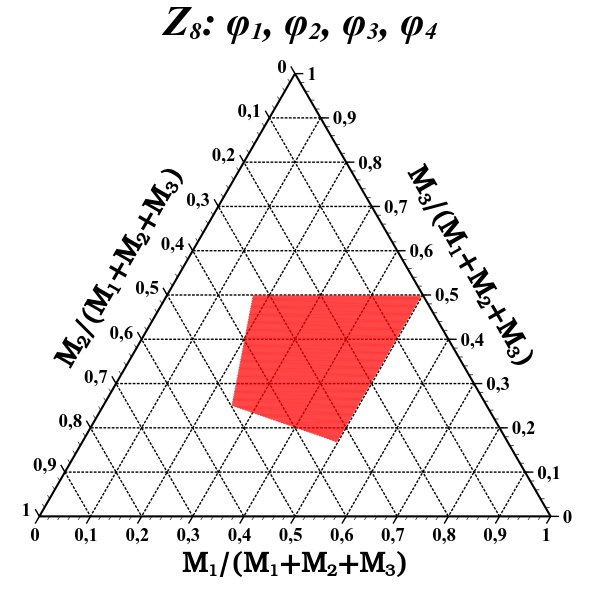}
\includegraphics[scale=0.5]{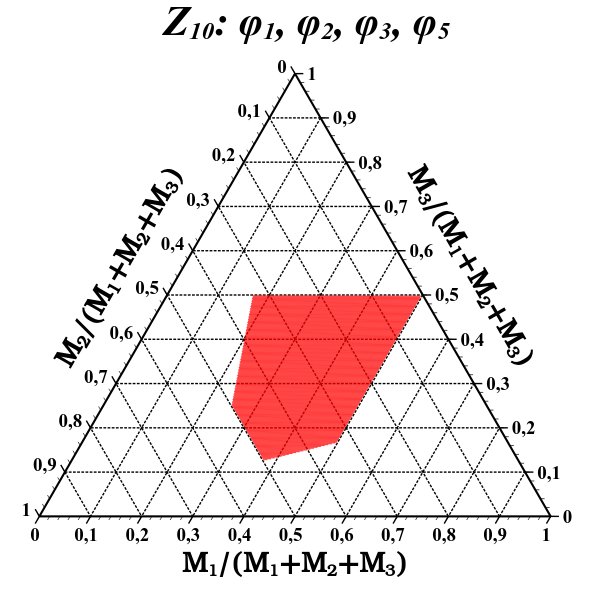}
\caption{The region where four fields can be stable for a  $Z_8$ symmetry with fields $\phi_1,\phi_2,\phi_3,\phi_4$ (left) and for  a  $Z_{10}$ symmetry with fields $\phi_1,\phi_2,\phi_3,\phi_5$ (right). For each point inside the colored region, it is possible to find values of $M'_4$ (left) or $M'_5$ (right) such that all four  fields are stable.  }
\label{fig:trianglez8fourstable}
\end{figure}

  The only possibility is $(\phi_1,\phi_2,\phi_3,\phi_4)$ with interaction terms given by
\begin{align}
 \mathcal{L}_3\supset&\,\, \phi _2 \phi _3^2,\phi _2^2 \phi _4,\phi _1 \phi _3 \phi _4,\phi _1^2 \phi _{\text{s2}},\phi
   _3 \phi _{\text{s1}} \phi _{\text{s2}},\phi _4 \phi _{\text{s2}}^2,\phi _4 \phi
   _{\text{s1}} \phi _{\text{s3}}.\\
  \mathcal{L}_4\supset&\,\, \phi _2^4,\phi _1 \phi _2^2 \phi _3,\phi _1^2 \phi _3^2,\phi _1^2 \phi _2 \phi _4,\phi _4^4,\phi _3^3 \phi
   _{\text{s1}},\phi _2 \phi _3 \phi _4 \phi _{\text{s1}},\phi _1 \phi _4^2 \phi _{\text{s1}},\phi _1^2 \phi _{\text{s1}}^2,\phi
   _3 \phi _{\text{s1}}^3,\phi _3^2 \phi _4 \phi _{\text{s2}},\phi _2 \phi _4^2 \phi _{\text{s2}},\nonumber\\
   &\phi _1 \phi _2 \phi
   _{\text{s1}} \phi _{\text{s2}},\phi _4 \phi _{\text{s1}}^2 \phi _{\text{s2}},\phi _2^2 \phi _{\text{s2}}^2,\phi _1 \phi _3
   \phi _{\text{s2}}^2,\phi _3 \phi _4^2 \phi _{\text{s3}},\phi _1 \phi _3 \phi _{\text{s1}} \phi _{\text{s3}},\phi _2 \phi _3 \phi _{\text{s2}} \phi
   _{\text{s3}},\phi _1 \phi _4 \phi _{\text{s2}} \phi _{\text{s3}},\nonumber\\
   &\phi
   _3^2 \phi _{\text{s3}}^2,\phi _2 \phi _4 \phi _{\text{s3}}^2,\phi _4^3 \phi _{\text{s4}},\phi _1 \phi _4 \phi _{\text{s1}} \phi _{\text{s4}},\phi _2 \phi
   _4 \phi _{\text{s2}} \phi _{\text{s4}},\phi _3 \phi _4 \phi
   _{\text{s3}} \phi _{\text{s4}},\phi _4^2 \phi _{\text{s4}}^2.\\
  \mathcal{L}_5\supset&\,\, \phi _1^3 \phi _2 \phi _3,\phi _1^4 \phi _4,\phi _3^4 \phi _4,\phi _2 \phi _3^2
   \phi _4^2,\phi _1 \phi _3 \phi _4^3,\phi _2^3 \phi _3 \phi _{\text{s1}},\phi _3^2
   \phi _4 \phi _{\text{s1}}^2,\phi _2 \phi _4^2 \phi _{\text{s1}}^2,\phi _4 \phi
   _{\text{s1}}^4,\phi _1 \phi _3^3 \phi _{\text{s2}},\nonumber\\
   &\phi _3 \phi _4^2 \phi _{\text{s1}} \phi _{\text{s2}},\phi _1 \phi _2
   \phi _4^2 \phi _{\text{s3}},\phi _4^3 \phi _{\text{s1}} \phi _{\text{s3}},\phi _4^2 \phi _{\text{s2}} \phi
   _{\text{s3}}^2,\phi _4 \phi
   _{\text{s3}}^4,\phi _3^2 \phi
   _{\text{s1}}^2 \phi _{\text{s4}} .
 \end{align}
 It follows that the full stability region is ensured by the condition 
 $M_2<2 M_3,\, M'_4<2 M_2,\, M_2<2 M_1,\, M_1<M_3+M'_4,\, M_3<M_1+M'_4,\,
   M'_4<M_1+M_3,\, M_1<M_2+M_3,\, M_3<M_1+M_2,\, M'_4<3 M_1$,
 %$M_2<\frac{M_1}{7}\land M_1-M_2<M_3\leq M_1\land M_1-M_3<M'_4<2 M_2$,
   and corresponds to the red region shown in Fig.~\ref{fig:trianglez8fourstable}. Inside that region it is possible to find values of $M'_4$ (not shown) such that all four particles are stable.

\subsection{$Z_{10}$}
With a $Z_{10}$ one can have up to five different fields. For concreteness, we will limit ourselves to the cases with more than three fields.
\subsubsection{Four fields}
There are four  possible sets of fields containing $\phi_5$, which we examine one by one:
\begin{itemize}
    \item $(\phi_1,\phi_2,\phi_3,\phi_5)$:  The interaction terms are
\begin{align}
 \mathcal{L}_3\supset&\,\, \phi _2 \phi _3 \phi _5,\phi _1^2 \phi
   _{\text{s2}},\phi _1 \phi _2 \phi
   _{\text{s3}},\phi _2 \phi _3 \phi
   _{\text{s5}}.\\
  \mathcal{L}_4\supset&\,\, \phi _2^2 \phi _3^2,\phi _1 \phi _3^3,\phi _1 \phi _2^2 \phi _5,\phi _1^2 \phi _3
   \phi _5,\phi _5^4,\phi _3^2 \phi _5 \phi _{\text{s1}},\phi _1 \phi _5^2 \phi
   _{\text{s1}},\phi _1^2 \phi _{\text{s1}}^2,\phi _3 \phi _{\text{s1}}^3,\phi _2 \phi
   _5^2 \phi _{\text{s2}},\phi _1 \phi _2 \phi _{\text{s1}} \phi _{\text{s2}},\nonumber\\
   &\phi _2^2
   \phi _{\text{s2}}^2,\phi _1 \phi _3 \phi _{\text{s2}}^2,\phi _5 \phi _{\text{s1}}
   \phi _{\text{s2}}^2,\phi _3 \phi _5^2 \phi
   _{\text{s3}},\phi _1 \phi _3 \phi
   _{\text{s1}} \phi _{\text{s3}},\phi _5 \phi _{\text{s1}}^2 \phi _{\text{s3}},\phi _2
   \phi _3 \phi _{\text{s2}} \phi _{\text{s3}},\phi _3^2 \phi _{\text{s3}}^2,\nonumber\\
   &\phi _1
   \phi _5 \phi _{\text{s3}}^2,\phi _5^3 \phi _{\text{s5}},\phi _1 \phi _5 \phi _{\text{s1}} \phi _{\text{s5}},\phi _2 \phi _5 \phi
   _{\text{s2}} \phi _{\text{s5}},\phi _3 \phi _5 \phi _{\text{s3}} \phi _{\text{s5}}.\\
   \mathcal{L}_5\supset&\,\, \phi _1 \phi _2^3 \phi _3,\phi _1^2 \phi _2 \phi _3^2,\phi _1^3 \phi _2 \phi
   _5,\phi _2 \phi _3 \phi _5^3,\phi _2 \phi _3^3 \phi _{\text{s1}},\phi _2^3 \phi _5
   \phi _{\text{s1}},\phi _2 \phi _5^2 \phi _{\text{s1}}^2,\phi _3^4 \phi
   _{\text{s2}},\phi _1 \phi _3^2 \phi _5 \phi _{\text{s2}},\nonumber\\
   &\phi _1^2 \phi _5^2 \phi
   _{\text{s2}},\phi _3 \phi _5^2 \phi _{\text{s1}} \phi _{\text{s2}},\phi _5 \phi
   _{\text{s1}}^3 \phi _{\text{s2}},\phi _1 \phi _5 \phi _{\text{s2}}^3,\phi _1 \phi _2
   \phi _5^2 \phi _{\text{s3}},\phi _5^3 \phi _{\text{s2}} \phi _{\text{s3}},\phi _2 \phi _5 \phi _{\text{s1}}
   \phi _{\text{s3}}^2.
 \end{align}
The stability condition is $M_2<M_3+M'_5,\, M_3<M_2+M'_5,\, M'_5<M_2+M_3,\, M_2<2 M_1,\, M_3<M_1+M_2,\,M_1<M_2+M_3,\, M_2<M_1+M_3,\, M_1<3 M_3$. This region is illustrated in the right panel of figure \ref{fig:trianglez8fourstable}. As before, inside the red region there exists values of $M'_5$ such that all four particles are stable.

\item $(\phi_1,\phi_2,\phi_4,\phi_5)$:  The interaction terms are
\begin{align}
 \mathcal{L}_3\supset&\,\, \phi _2 \phi _4^2,\phi _1 \phi _4 \phi _5,\phi _1^2
   \phi _{\text{s2}},\phi _2^2 \phi _{\text{s4}},\phi _1
   \phi _4 \phi _{\text{s5}}.\\
  \mathcal{L}_4\supset&\,\, \phi _2^3 \phi _4,\phi _1^2 \phi _4^2,\phi _1 \phi _2^2 \phi _5,\phi _5^4,\phi _2
   \phi _4 \phi _5 \phi _{\text{s1}},\phi _1 \phi _5^2 \phi _{\text{s1}},\phi _1^2 \phi
   _{\text{s1}}^2,\phi _4^3 \phi _{\text{s2}},\phi _2 \phi _5^2 \phi _{\text{s2}},\phi
   _1 \phi _2 \phi _{\text{s1}} \phi _{\text{s2}},\phi _4 \phi _{\text{s1}}^2 \phi
   _{\text{s2}},\nonumber\\
   &\phi _2^2 \phi _{\text{s2}}^2,\phi _5 \phi _{\text{s1}} \phi
   _{\text{s2}}^2,\phi _4 \phi _5^2 \phi
   _{\text{s4}},\phi _1 \phi _4 \phi _{\text{s1}} \phi _{\text{s4}},\phi _2 \phi _4 \phi
   _{\text{s2}} \phi _{\text{s4}},\phi _1 \phi _5 \phi _{\text{s2}} \phi
   _{\text{s4}},\phi _4^2 \phi _{\text{s4}}^2,\phi _5^3 \phi _{\text{s5}},\phi _1 \phi _5 \phi _{\text{s1}} \phi _{\text{s5}},\nonumber\\
   &\phi _2 \phi _5 \phi
   _{\text{s2}} \phi _{\text{s5}},\phi _4 \phi _5 \phi _{\text{s4}} \phi _{\text{s5}},\phi _5^2 \phi _{\text{s5}}^2.\\
   \mathcal{L}_5\supset&\,\, \phi _1^2 \phi _2^2 \phi _4,\phi _1^3 \phi _2 \phi _5,\phi _2 \phi _4^2 \phi
   _5^2,\phi _1 \phi _4 \phi _5^3,\phi _2^3 \phi _5 \phi _{\text{s1}},\phi _2 \phi _5^2
   \phi _{\text{s1}}^2,\phi _4 \phi _{\text{s1}}^4,\phi _1^2 \phi _5^2 \phi
   _{\text{s2}},\phi _4^2 \phi _5 \phi _{\text{s1}} \phi _{\text{s2}},\phi _5 \phi
   _{\text{s1}}^3 \phi _{\text{s2}},\nonumber\\
   &\phi _4 \phi _5^2 \phi _{\text{s2}}^2,\phi _1 \phi
   _5 \phi _{\text{s2}}^3,\phi _2^2 \phi _5^2 \phi
   _{\text{s4}},\phi _5^3 \phi _{\text{s1}} \phi _{\text{s4}},\phi _1 \phi _2 \phi _5 \phi _{\text{s4}}^2,\phi
   _5^2 \phi _{\text{s2}} \phi _{\text{s4}}^2.
 \end{align}
 The stability condition reads $M_2<2 M_1,\, M_4<2 M_2,\, M_2<2 M_4,\, M_1<M_4+M'_5,\, M_4<M_1+M'_5,\,M'_5<M_1+M_4$. 
   
    \item $(\phi_1,\phi_3,\phi_4,\phi_5)$:  The interaction terms are
\begin{align}
 \mathcal{L}_3\supset&\,\,\phi _3^2 \phi _4,\phi _1 \phi _4 \phi _5,\phi _1 \phi _3 \phi _{\text{s4}},\phi _1 \phi _4 \phi
   _{\text{s5}}.\\
  \mathcal{L}_4\supset&\,\, \phi _1 \phi _3^3,\phi _1^2 \phi _4^2,\phi _1^2 \phi _3 \phi _5,\phi _5^4,\phi _3
   \phi _4^2 \phi _{\text{s1}},\phi _3^2 \phi _5 \phi _{\text{s1}},\phi _1 \phi _5^2
   \phi _{\text{s1}},\phi _1^2 \phi _{\text{s1}}^2,\phi _1^3
   \phi _{\text{s3}},\phi _4^2 \phi _5 \phi _{\text{s3}},\phi _3 \phi _5^2 \phi
   _{\text{s3}},\nonumber\\
   &\phi _1 \phi _3 \phi _{\text{s1}} \phi _{\text{s3}},\phi _5 \phi
   _{\text{s1}}^2 \phi _{\text{s3}},\phi _3^2 \phi _{\text{s3}}^2,\phi _1 \phi _5 \phi
   _{\text{s3}}^2,\phi _4 \phi _5^2 \phi
   _{\text{s4}},\phi _1 \phi _4 \phi _{\text{s1}} \phi _{\text{s4}},\phi _3 \phi _4 \phi
   _{\text{s3}} \phi _{\text{s4}},\phi _4^2 \phi _{\text{s4}}^2,\nonumber\\
   &\phi _3 \phi _5 \phi
   _{\text{s4}}^2,\phi _5^3 \phi _{\text{s5}},\phi _1 \phi _5 \phi _{\text{s1}} \phi
   _{\text{s5}},\phi _3 \phi _5 \phi
   _{\text{s3}} \phi _{\text{s5}},\phi _4 \phi _5 \phi
   _{\text{s4}} \phi _{\text{s5}}.\\
   \mathcal{L}_5\supset&\,\, \phi _1^3 \phi _3 \phi _4,\phi _3 \phi _4^3 \phi _5,\phi _3^2 \phi _4 \phi
   _5^2,\phi _1 \phi _4 \phi _5^3,\phi _3 \phi _4 \phi _5 \phi _{\text{s1}}^2,\phi _4
   \phi _{\text{s1}}^4,\phi _1 \phi _4^3 \phi _{\text{s3}},\phi _4 \phi _5^2 \phi
   _{\text{s1}} \phi _{\text{s3}},\phi _1^2 \phi _4 \phi _{\text{s3}}^2,\nonumber\\
   &\phi _4 \phi _5
   \phi _{\text{s3}}^3,\phi _3^3 \phi _5 \phi
   _{\text{s4}},\phi _1 \phi _3 \phi _5^2 \phi _{\text{s4}},\phi _5^3 \phi _{\text{s1}}
   \phi _{\text{s4}},\phi _1^2 \phi _5
   \phi _{\text{s3}} \phi _{\text{s4}},\phi _5^2 \phi _{\text{s3}}^2 \phi _{\text{s4}},\phi _5 \phi _{\text{s3}} \phi _{\text{s4}}^3.
 \end{align}
 The stability condition is $M_4<2 M_3,\, M_1<M_3+M_4,\, M_3<M_1+M_4,\, M_4<M_1+M_3,\,M_1<M_4+M'_5,\,M_4<M_1+M'_5,\,M'_5<M_1+M_4,\,M_3<3 M_1$. 
\item $(\phi_2,\phi_3,\phi_4,\phi_5)$:  The interaction terms are
\begin{align}
 \mathcal{L}_3\supset&\,\, \phi _3^2 \phi _4,\phi _2 \phi _4^2,\phi _2 \phi _3 \phi _5,\phi _4 \phi
   _{\text{s2}}^2,\phi _2 \phi _3 \phi _{\text{s5}}.\\
  \mathcal{L}_4\supset&\,\, \phi _2^2 \phi _3^2,\phi _2^3 \phi _4,\phi _5^4,\phi _4^3 \phi _{\text{s2}},\phi
   _3 \phi _4 \phi _5 \phi _{\text{s2}},\phi _2 \phi _5^2 \phi _{\text{s2}},\phi _2^2
   \phi _{\text{s2}}^2,\phi _4^2 \phi _5 \phi _{\text{s3}},\phi _3 \phi _5^2 \phi
   _{\text{s3}},\phi _2 \phi _3 \phi _{\text{s2}} \phi _{\text{s3}},\phi _3^2 \phi
   _{\text{s3}}^2,\nonumber\\
   &\phi _2 \phi _4 \phi _{\text{s3}}^2,\phi _4 \phi _5^2 \phi _{\text{s4}},\phi _2 \phi _4 \phi _{\text{s2}} \phi _{\text{s4}},\phi _3 \phi _4 \phi _{\text{s3}} \phi _{\text{s4}},\phi _2 \phi _5
   \phi _{\text{s3}} \phi _{\text{s4}},\phi _4^2 \phi _{\text{s4}}^2,\phi _3 \phi _5
   \phi _{\text{s4}}^2,\phi _5^3 \phi _{\text{s5}},\phi _2 \phi _5 \phi _{\text{s2}} \phi
   _{\text{s5}},\nonumber\\
   &\phi _3 \phi _5 \phi
   _{\text{s3}} \phi _{\text{s5}},\phi _4 \phi _5 \phi _{\text{s4}} \phi
   _{\text{s5}},\phi _5^2 \phi _{\text{s5}}^2.\\
   \mathcal{L}_5\supset&\,\, \phi _3 \phi _4^3 \phi _5,\phi _3^2 \phi _4 \phi _5^2,\phi _2 \phi _4^2 \phi
   _5^2,\phi _2 \phi _3 \phi _5^3,\phi _3^4 \phi _{\text{s2}},\phi _4 \phi _5^2 \phi
   _{\text{s2}}^2,\phi _2^2 \phi _4 \phi _5 \phi _{\text{s3}},\phi _5^3 \phi
   _{\text{s2}} \phi _{\text{s3}},\phi _4^2 \phi _{\text{s2}} \phi _{\text{s3}}^2,\phi
   _4 \phi _5 \phi _{\text{s3}}^3,\nonumber\\
   &\phi _3^3 \phi _5 \phi
   _{\text{s4}},\phi _2^2 \phi _5^2 \phi _{\text{s4}},\phi _3 \phi _5 \phi
   _{\text{s2}}^2 \phi _{\text{s4}},\phi _5^2 \phi _{\text{s3}}^2 \phi _{\text{s4}},\phi _5^2 \phi _{\text{s2}} \phi _{\text{s4}}^2,\phi
   _5 \phi _{\text{s3}} \phi _{\text{s4}}^3.
 \end{align}
 The stability condition reads $M_4<2 M_2,\, M_4<2 M_3,\, M_2<2 M_4,\,M_2<M_3+M'_5.\, M_3<M_2+M'_5,\,M'_5<M_2+M_3$. 
 
\end{itemize}
\subsubsection{Five fields}
There is just one choice for the fields, $(\phi_1,\ldots,\phi_5)$, with interaction terms given by
    \begin{align}
 \mathcal{L}_3\supset&\,\, \phi _3^2 \phi _4,\phi _2 \phi _4^2,\phi _2 \phi _3 \phi _5,\phi _1 \phi _4 \phi
   _5,\phi _1^2 \phi _{\text{s2}},\phi _3 \phi _{\text{s1}}
   \phi _{\text{s2}},\phi _4 \phi _{\text{s2}}^2,\phi
   _4 \phi _{\text{s1}} \phi _{\text{s3}},\phi _5 \phi _{\text{s2}} \phi
   _{\text{s3}},\phi _1 \phi _4\phi _{\text{s5}}.
   \end{align}
   \begin{align}
  \mathcal{L}_4\supset&\,\, \phi _2^2 \phi _3^2,\phi _1 \phi _3^3,\phi _2^3 \phi _4,\phi _1 \phi _2 \phi _3
   \phi _4,\phi _1^2 \phi _4^2,\phi _1 \phi _2^2 \phi _5,\phi _1^2 \phi _3 \phi _5,\phi
   _5^4,\phi _3 \phi _4^2 \phi _{\text{s1}},\phi _3^2 \phi _5 \phi _{\text{s1}},\phi _2
   \phi _4 \phi _5 \phi _{\text{s1}},\nonumber\\
   &\phi _1 \phi _5^2 \phi _{\text{s1}},\phi _1^2 \phi
   _{\text{s1}}^2,\phi _3 \phi _{\text{s1}}^3,\phi _4^3 \phi _{\text{s2}},\phi _3 \phi
   _4 \phi _5 \phi _{\text{s2}},\phi _2 \phi _5^2 \phi _{\text{s2}},\phi _1 \phi _2 \phi
   _{\text{s1}} \phi _{\text{s2}},\phi _4 \phi _{\text{s1}}^2 \phi _{\text{s2}},\phi
   _2^2 \phi _{\text{s2}}^2,\phi _1 \phi _3 \phi _{\text{s2}}^2,\nonumber\\
   &\phi _5 \phi
   _{\text{s1}} \phi _{\text{s2}}^2,\phi _4^2 \phi _5 \phi
   _{\text{s3}},\phi _3 \phi _5^2 \phi _{\text{s3}},\phi _1 \phi _3 \phi _{\text{s1}} \phi _{\text{s3}},\phi _5 \phi
   _{\text{s1}}^2 \phi _{\text{s3}},\phi _2 \phi _3 \phi _{\text{s2}} \phi
   _{\text{s3}},\phi _1 \phi _4 \phi _{\text{s2}} \phi _{\text{s3}},\phi _3^2 \phi
   _{\text{s3}}^2,\nonumber\\
   &\phi _2 \phi _4 \phi _{\text{s3}}^2,\phi _1 \phi _5 \phi
   _{\text{s3}}^2,\phi _4 \phi _5^2 \phi
   _{\text{s4}},\phi _1 \phi _4 \phi
   _{\text{s1}} \phi _{\text{s4}},\phi _2
   \phi _4 \phi _{\text{s2}} \phi _{\text{s4}},\phi _1 \phi _5 \phi _{\text{s2}} \phi
   _{\text{s4}},\phi _3 \phi _4 \phi _{\text{s3}}
   \phi _{\text{s4}},\phi _2 \phi _5 \phi _{\text{s3}} \phi _{\text{s4}},\nonumber\\
   &\phi _4^2 \phi
   _{\text{s4}}^2,\phi _3 \phi _5 \phi _{\text{s4}}^2,\phi _5^3 \phi _{\text{s5}},\phi _1 \phi _5 \phi
   _{\text{s1}} \phi _{\text{s5}},\phi _2 \phi _5 \phi _{\text{s2}} \phi _{\text{s5}},\phi _3 \phi _5 \phi _{\text{s3}} \phi _{\text{s5}},\phi _4 \phi _5 \phi _{\text{s4}} \phi _{\text{s5}},\phi
   _5^2 \phi _{\text{s5}}^2.
   %%%%%%%%%%%%%%%%%%%%%%%%%%%%%%%%%%%%%%%%%%%%%%%%%%%%%%%%%%%%%%%%%%%%%%%%%
    \end{align}
   \begin{align}
   \mathcal{L}_5\supset&\,\, \phi _1 \phi _2^3 \phi _3,\phi _1^2 \phi _2 \phi _3^2,\phi _1^2 \phi _2^2 \phi
   _4,\phi _1^3 \phi _3 \phi _4,\phi _1^3 \phi _2 \phi _5,\phi _3 \phi _4^3 \phi _5,\phi
   _3^2 \phi _4 \phi _5^2,\phi _2 \phi _4^2 \phi _5^2,\phi _2 \phi _3 \phi _5^3,\phi _1
   \phi _4 \phi _5^3,\nonumber\\
   &\phi _2 \phi _3^3 \phi _{\text{s1}},\phi _2^2 \phi _3 \phi _4 \phi
   _{\text{s1}},\phi _2^3 \phi _5 \phi _{\text{s1}},\phi _3 \phi _4 \phi _5 \phi
   _{\text{s1}}^2,\phi _2 \phi _5^2 \phi _{\text{s1}}^2,\phi _4 \phi _{\text{s1}}^4,\phi
   _3^4 \phi _{\text{s2}},\phi _1 \phi _3 \phi _4^2 \phi _{\text{s2}},\phi _1 \phi _3^2
   \phi _5 \phi _{\text{s2}},\nonumber\\
   &\phi _1^2 \phi _5^2 \phi _{\text{s2}},\phi _4^2 \phi _5
   \phi _{\text{s1}} \phi _{\text{s2}},\phi _3 \phi _5^2 \phi _{\text{s1}} \phi
   _{\text{s2}},\phi _5 \phi _{\text{s1}}^3 \phi _{\text{s2}},\phi _4 \phi _5^2 \phi
   _{\text{s2}}^2,\phi _1 \phi _5 \phi _{\text{s2}}^3,\phi _1 \phi _4^3 \phi
   _{\text{s3}},\phi _2^2 \phi _4 \phi _5 \phi _{\text{s3}},\nonumber\\
   &\phi _1 \phi _2 \phi _5^2
   \phi _{\text{s3}},\phi _4 \phi _5^2 \phi _{\text{s1}} \phi _{\text{s3}},\phi _5^3
   \phi _{\text{s2}} \phi _{\text{s3}},\phi _1^2 \phi _4 \phi _{\text{s3}}^2,\phi _2 \phi _5 \phi _{\text{s1}}
   \phi _{\text{s3}}^2,\phi _4^2 \phi _{\text{s2}} \phi _{\text{s3}}^2,\phi _4 \phi _5 \phi
   _{\text{s3}}^3,\phi _3^3 \phi _5 \phi _{\text{s4}},\nonumber\\
   &\phi
   _2^2 \phi _5^2 \phi _{\text{s4}},\phi _1 \phi _3 \phi _5^2 \phi _{\text{s4}},\phi
   _5^3 \phi _{\text{s1}} \phi _{\text{s4}},\phi _3 \phi _5 \phi _{\text{s2}}^2 \phi _{\text{s4}},\phi _1^2 \phi _5 \phi
   _{\text{s3}} \phi _{\text{s4}},\phi
   _5^2 \phi _{\text{s3}}^2 \phi _{\text{s4}},\phi
   _1 \phi _2 \phi _5 \phi _{\text{s4}}^2,\nonumber\\
   &\phi _5^2 \phi _{\text{s2}} \phi
   _{\text{s4}}^2,\phi _5 \phi _{\text{s3}} \phi
   _{\text{s4}}^3. 
 \end{align}
 The stability condition for the five fields is $M_2<2 M_1,\, M_1<M_2+M_3,\, M_2<M_1+M_3,\, M_3<M_1+M_2,\, M_4<2 M_2,\,
   M_1<M_3+M_4,\, M_3<M_1+M_4,\, M_4<M_1+M_3,\, M_4<2 M_3,\, M_2<2 M_4,\,
   M_2<M_3+M'_5,\, M_3<M_2+M'_5,\, M'_5<M_2+M_3,\, M_1<M_4+M'_5,\, M_4<M_1+M'_5,\,M'_5<M_1+M_4$.

\section{Discussion}
\label{sec:disc}

Let us first summarize the main results found in the previous two sections regarding multi-component dark matter scenarios under different $Z_N$ symmetries:
\begin{itemize}
    \item $Z_4$. This is the smallest $Z_N$ symmetry that allows a two-component dark matter scenario. Only one realization is possible, in which the dark matter consists of a real scalar field ($\phi'_2$) and a complex scalar field ($\phi_1$).  
    \item $Z_5$.  A unique realization of two-component dark matter is possible, with both  DM particles $(\phi_1,\phi_2)$ being complex. 
    \item $Z_6$. This is the smallest $Z_N$ symmetry that leads to a scenario with three dark matter particles: two complex scalar fields ($\phi_{1,2}$) and one real scalar field ($\phi'_3$). It   also provides the simplest example of  \emph{unconditional stability}  for two dark matter particles (one real and one complex). Three different realizations of two-component dark matter are possible. 
    \item $Z_7$. Here all the scenarios involve complex DM particles, and both two and three DM particles are allowed. It is the smallest $Z_N$ symmetry for  which the DM may consists of three complex scalars. 
    \item $Z_8$.  It is the smallest  $Z_N$ symmetry that leads to  a four-component dark matter scenario. One of those particles is real while the other three are complex. Two, three or four DM particles can be obtained within this symmetry. Unconditional stability for two particles appears in two scenarios but limited to the renormalizable level.   
    \item $Z_9$. The DM particles are all complex and there may be  up to four of them. Several  scenarios with two and three dark matter particles can be envisaged.
    \item $Z_{10}$. Up to five DM particles can be realized within this symmetry, with complex DM for those cases not considering $\phi_5$.   
\end{itemize}

These results clearly indicate that there is plenty of viable and interesting scenarios to explore for multi-component dark matter under a $Z_N$ symmetry. All of them feature several particles (scalar fields) that are stable and have the right particle-physics properties to account for (a fraction of) the observed dark matter density. If these particles are going to actually explain the dark matter, we must ensure, in addition,  that their relic density is consistent with the observations and that they satisfy current experimental limits --mainly  from direct and indirect dark matter experiments but also from collider searches. So far, this analysis has not been done for \emph{any} of the models outlined in this work. Even though a detailed study of these issues lies beyond the scope of the present paper, some generic features can be briefly described. 

\begin{figure}[t]
\centering
\includegraphics[scale=0.9]{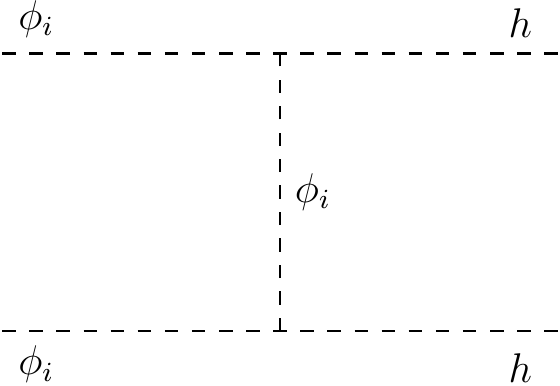}\hspace{1cm}
\includegraphics[scale=0.9]{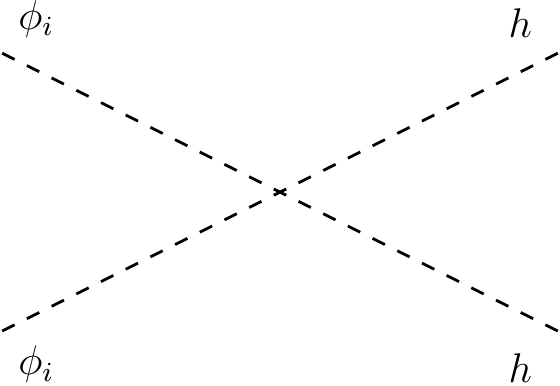}\\

\vspace{0.3cm}
\includegraphics[scale=0.9]{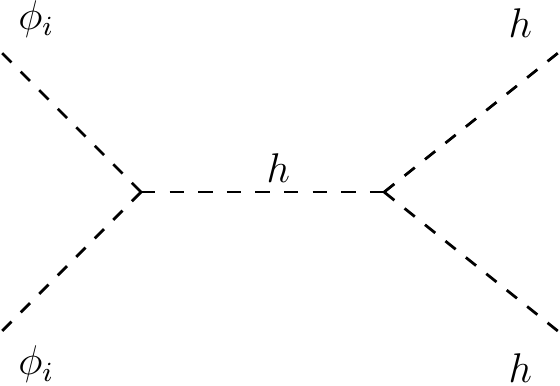}\hspace{1cm}
\includegraphics[scale=0.9]{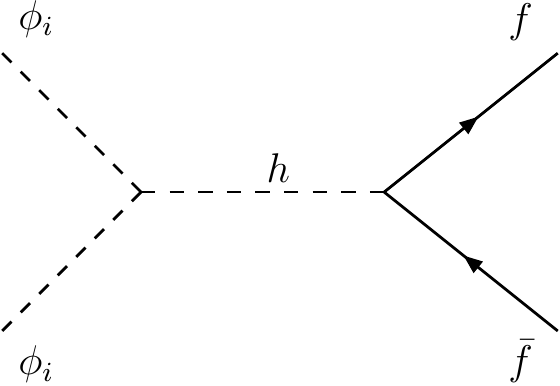}
\caption{DM annihilation channels via cubic and quartic interactions with the Higgs. Notice that one more diagram is present by replacing the SM fermions with the SM gauge bosons.}
\label{fig:annihilationtoSM}
\end{figure}
\begin{figure}
\centering
\includegraphics[scale=0.9]{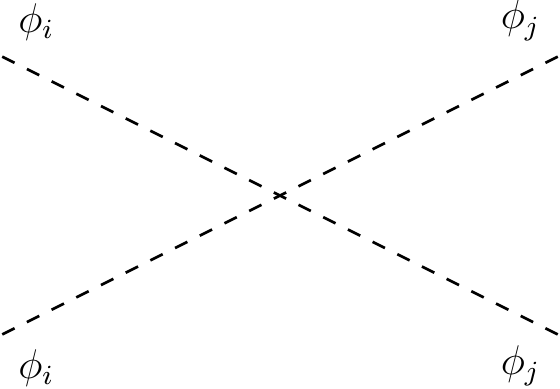}\hspace{1cm}
\includegraphics[scale=0.9]{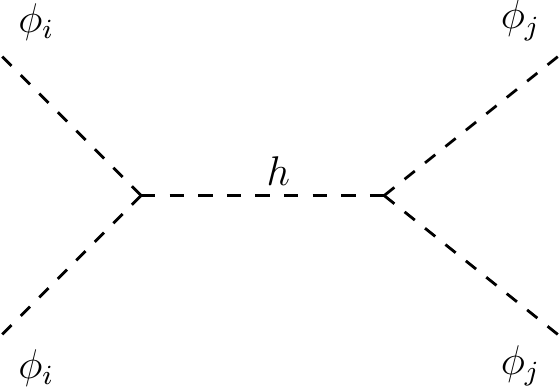}
\caption{$\phi_i\phi_i\leftrightarrow\phi_j\phi_j$ DM conversion channels.}
\label{fig:conversion1}
\end{figure}
\begin{figure}
\centering
\includegraphics[scale=0.9]{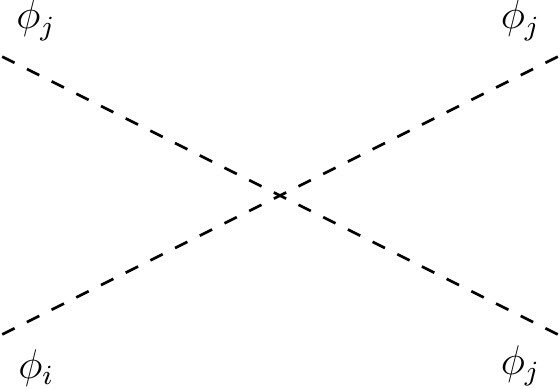}\hspace{1cm}\includegraphics[scale=0.9]{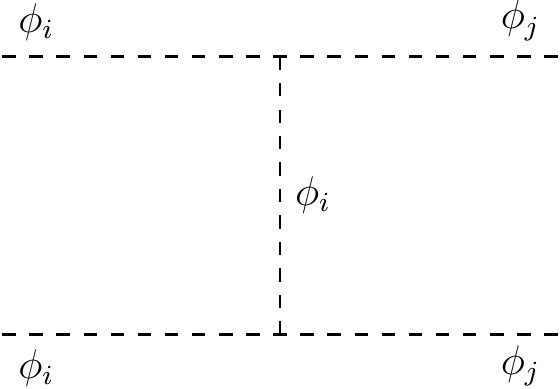}
\caption{$\phi_i\phi_j\to\phi_j\phi_j$ DM conversion channels via quartic (left panel) and trilinear (right panel) interactions involving a linear term on $\phi_{i}$. %Notice that may exist other diagrams with final states different to  $\phi_j$, {\it i.e.} $\phi_i\phi_j\to\phi_k\phi_l$.
}
%\label{fig:conversion2}
\label{fig:conversion2}
\end{figure}

\begin{figure}
\centering
\includegraphics[scale=0.9]{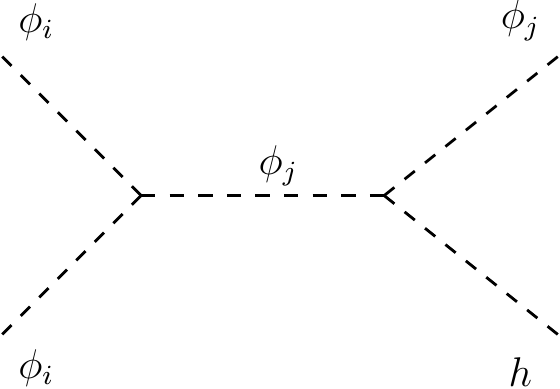}\hspace{1cm}
\includegraphics[scale=0.9]{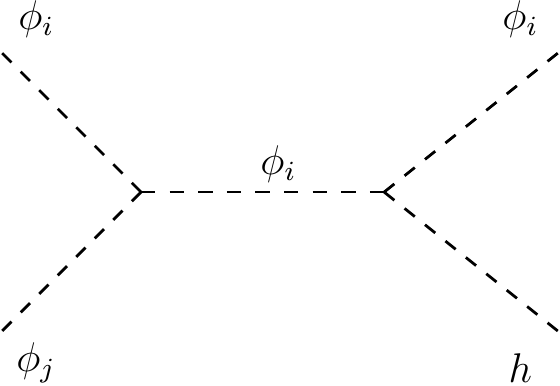}\hspace{1cm}
\includegraphics[scale=0.9]{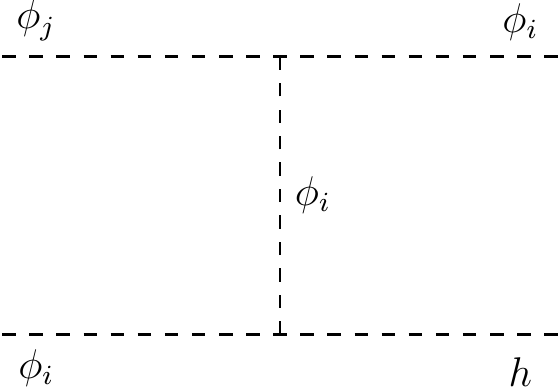}
\caption{$\phi_i,\phi_j$ DM semiannihilation channels via cubic interactions involving linear terms on $\phi_{i}$.}
\label{fig:semianni2}
\end{figure}

\begin{figure}
\centering
\includegraphics[scale=0.9]{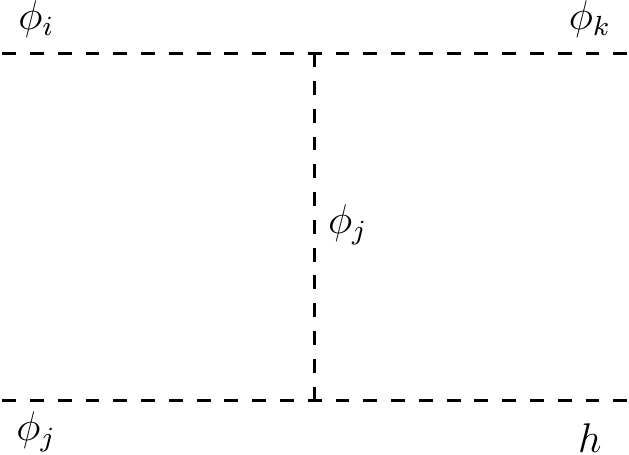}\hspace{1cm}
\includegraphics[scale=0.9]{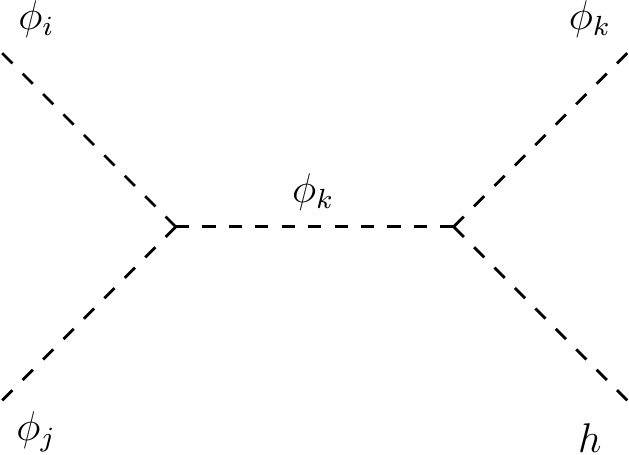}
\caption{DM semiannihilation channels involving three DM fields.}
\label{fig:semianni3}
\end{figure}

\begin{figure}
\centering
\includegraphics[scale=0.9]{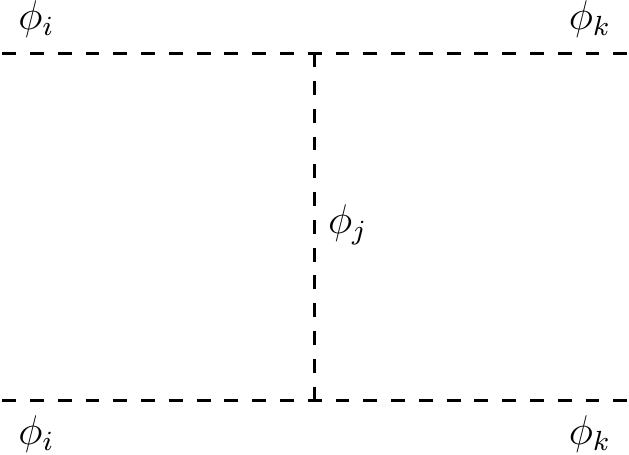}\hspace{1cm}
\includegraphics[scale=0.9]{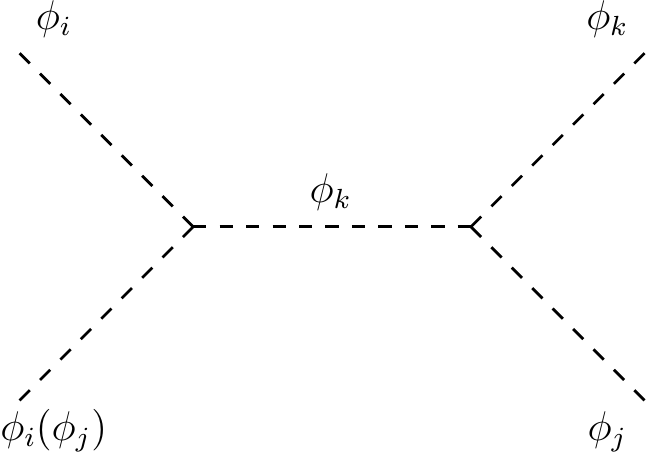}\\

\vspace{0.3cm}
\includegraphics[scale=0.9]{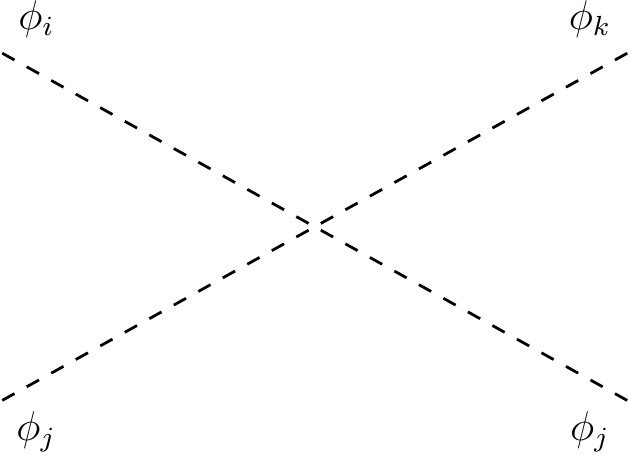}
\caption{DM conversion channels involving three DM fields.}
\label{fig:conversion3}
\end{figure}

\begin{figure}
\centering
\includegraphics[scale=0.9]{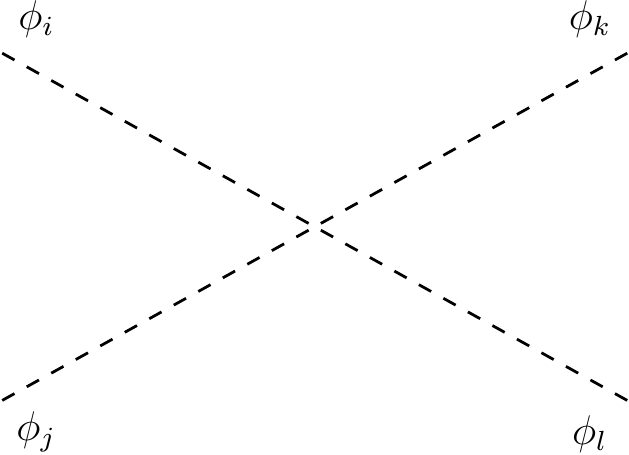}\hspace{1cm}
\includegraphics[scale=0.9]{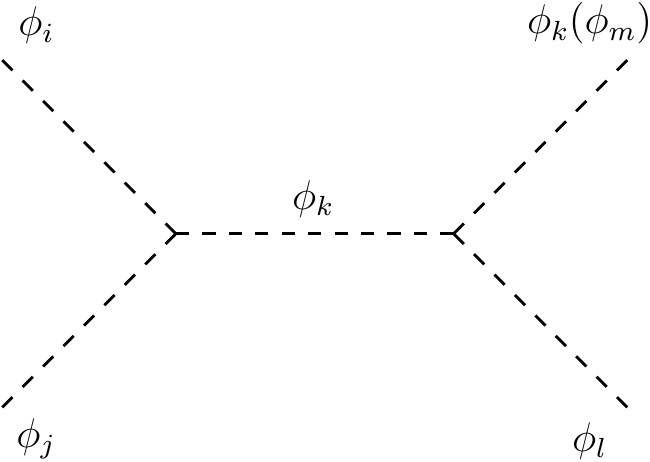}
\caption{DM conversion channels involving four or more DM fields. There exists also a $t$-channel diagram obtained from the $s$-channel diagram. }
\label{fig:conversion4}
\end{figure}

For the scenarios we are considering, where the only new particles are scalars that are SM singlets, the portal linking the dark and visible sectors is the interaction with the Higgs field, which is neutral under the $Z_N$ symmetry. Concretely, the only $Z_N$ invariant scalar interactions with the Higgs have the form $\lambda_{H\phi_i} H^\dagger H\phi_i^\dagger\phi_i$, 
thus leading to the DM annihilation into SM particles through the processes displayed in Fig.~\ref{fig:annihilationtoSM}. In contrast, co-annihilation processes \cite{Edsjo:1997bg} such $\phi_i\phi_j\to\, SM$ are forbidden since the mixing terms $\phi_i\phi_j$ and $\phi_i^\dagger\phi_j$ ($i\neq j$) are not allowed to take place in the Lagrangian. 
At first sight, it appears that each relic density $\Omega_{\phi_i}$ only depends on $M_i$ and the scalar coupling $\lambda_{H\phi_i}$ \cite{McDonald:1993ex} (only $\phi_i$ annihilations via the Higgs portal would be acting). If so, the current experimental limits would lead to two viable regions: one around $M_i\approx m_h/2$ and the other  one  at $M_i\gtrsim \mathcal{O}$(1) TeV~\cite{Cline:2013gha,Athron:2017kgt}.
Since the annihilation through $s$-channel Higgs boson exchange tends to dominate the total annihilation cross section, these viable mass regions would remain despite each  $\phi_i$ contributing  less than $100\%$ of the total DM abundance.

Nonetheless, the interplay between  all the scalar interactions may alter these results.  For instance, the combination of the interactions $H^\dagger H\phi_i^\dagger\phi_i $ and  $(\phi_i^\dagger\phi_i) (\phi_j^\dagger\phi_j)$ (or similar terms) gives rise to the DM conversion processes~\cite{Liu:2011aa,Adulpravitchai:2011ei,Belanger:2011ww} (see Fig.~\ref{fig:conversion1})  where the individual $\phi_i$ particle number changes but the total number of $\phi$'s particles remains constant.

Furthermore, the operators leading to two- and three-body decays of DM particles also generate DM conversion processes as those shown in the left panel of Fig.~\ref{fig:conversion2}, where the $\phi_i$ particle number changes in one unit\footnote{Notice that other number changing processes such as $3\to2$ DM annihilations may arise, which, under certain conditions \cite{Hochberg:2014dra}, would be relevant for the relic density calculation.}. And the interplay of the two-body decay terms with the DM-Higgs interactions allows for semi-annihilation processes \cite{Hambye:2008bq,DEramo:2010keq} such of those in Fig. \ref{fig:semianni2}.
All in all, it is expected that these additional processes may significantly modify~\cite{DEramo:2010keq,Belanger:2012vp} the typical outcome
for the relic density of each DM particle from the standard freeze-out process~\cite{Steigman:2012nb}.

%\textcolor{blue}{FEATURES WITH 3 OR MORE FIELDS.\\
A direct consequence of the existence of several DM fields charged under the same $Z_N$ is the presence in the Lagrangian of cubic and quartic interaction terms involving only one single DM fields, {\it e.g.} $\phi_i\phi_j^2$, $\phi_i\phi_j\phi_k$ and $\phi_i\phi_j\phi_k\phi_l$ with $i\neq j\neq k\neq l$ (notice that in DM frameworks with a direct product of $Z_N$ symmetries, such as $Z_2\otimes Z_2'$ or $Z_3\otimes Z_3'\otimes Z_3''$, that can not occur since each single field is only charged under the corresponding symmetry). 
These terms in turn lead to extra semi-annihilation and DM-conversion processes as those displayed in Figs. \ref{fig:semianni3}-\ref{fig:conversion4}, with the former playing a main role in the annihilation of the lightest DM particle (notice that its coupling to the Higgs can be arbitrarily small).   %(notice that the resonant semi-annihilation \cite{Bhattacharya:2017fid} is forbidden by the stability conditions). 
Furthermore, in case of a small interaction between the Higgs and the lightest DM candidate, say $\lambda_{S1}\phi_1^*\phi_1 h\supset\lambda_{S1}\phi_1^*\phi_1 |H|^2$, the other DM particles may generate at one-loop level such a interaction, for instance through a triangle loop with a single $\phi_i$ ($i\neq 1$) running in the loop, thus softening the direct detection constraints on $\phi_1$ \cite{Casas:2017jjg}. 

To illustrate the novel features of the framework we are presenting, let us consider the scenario with three complex scalar fields $(\phi_1,\phi_2,\phi_3)$ charged under a $Z_7$ symmetry,  
\begin{align}
    \phi_1\sim \omega_7,\,\,\, \phi_2\sim \omega_7^2,\,\,\, \phi_3\sim \omega_7^3; \hspace{1cm}\omega_7=\exp(i2\pi/7). 
\end{align}
It follows that the most general $Z_7$-invariant scalar potential include the following interactions:
\begin{align}\label{eq:LZ7}
 \mathcal{V}&\supset\,\,\lambda_{Si}|H|^2|\phi_i|^2+\lambda_{4ij}|\phi_i|^2|\phi_j|^2+\left[\mu_{1}\phi _2^2 \phi _3+\mu_{2}\phi_1 \phi_3^2+\mu_{3}\phi_1^2 \phi_2^*+\mu_{4}\phi_1\phi_2\phi_3^*+ \text{H.c.}\right]\nonumber\\
 & \,+\left[\lambda_1 \phi _1 \phi _2^3+\lambda_2\phi _1^2 \phi _2 \phi _3+\lambda_3\phi _2 \phi _3^2 \phi_1^*+\lambda_4\phi _3\phi_1^{*3}+\lambda_5\phi _3^3 \phi_2^*+\lambda_6\phi_1\phi_3\phi_2^{*2}+ \text{H.c.}\right],
 \end{align}
 where $H$ is the SM Higgs boson. 
All the trilinear and the quartic interactions in brackets are new in comparison to scenarios with several discrete symmetries. The quartic interactions mediate DM conversion processes while the trilinear ones mediate both DM conversion and semi-annihilation (along with $\lambda_{Si}$) processes.
%

%ADDITIONAL FEATURES.\\
When one of some of the $\phi$ fields are not stable, then they only may decay into DM fields. That is, the decays into the visible sector such as $\phi_i\to\phi_i+h^{(*)}\to\phi_i+\gamma+\gamma$ are forbidden.  Thus the detection of an indirect signal of this class \cite{Herms:2019mnu,Ghosh:2019jzu} would rule out our framework. 

%} 

Regarding direct (DD) and indirect (ID) DM searches, each singlet can  scatter elastically on nuclei and self-annihilate as in the one-component DM scenario, {\it i.e.}, via $t$-channel and $s$-channel Higgs boson exchange, respectively. 
Nevertheless, the alteration of the standard DM freeze-out process due to the existence of additional DM annihilation processes automatically affects the DM phenomenology  in comparison with the one-component DM scenario\footnote{It worth mentioning that multicomponent DM scenarios may lead to a weakening of the bounds on the corresponding rates since these depend on the local DM density $-$implying that a rescaling in the DD and ID observables associated to each DM particle should be taken into account. }.
Moreover, the semi-annihilation processes may also play a important role in ID searches due to the presence of new annihilation channels~\cite{DEramo:2010keq,DEramo:2012fou,Belanger:2014bga}. 
On the other hand, invisible Higgs decays $h\to \phi_i\phi_j$ are expected to occur if the DM particles are sufficiently light, in which case the LHC upper bound on the invisible branching ratio $\text{BR}_{inv}< 0.19(0.26)$ \cite{Sirunyan:2018owy,Aaboud:2019rtt} applies. 

Another appealing alternative to explain the relic density and satisfy current experimental limits is via \emph{freeze-in} \cite{Hall:2009bx,Bernal:2017kxu}.  If the Higgs portal couplings are all tiny, $\lambda_{H\phi_i}\ll 1$, the new scalars  would never reach thermal equilibrium in the early Universe, preventing  a freeze-out process. They  would still be slowly produced, though,  from  the decays and scatterings of the particles in the thermal plasma --a process dubbed freeze-in \cite{Klasen:2013ypa,Molinaro:2014lfa}. In this case, results similar to those found for the singlet scalar are expected \cite{Yaguna:2011qn}. Such tiny couplings would also guarantee that  the usual signals at colliders and at direct and indirect detection experiments remain unobservable. On the other hand, establishing that the dark matter actually consists of more than one particle would become significantly more challenging.

One may wonder if there exists any  advantages  in using a  $Z_N$ rather than other discrete symmetries to stabilize multiple  dark matter particles and whether it is possible to discriminate between these possibilities.  The Klein group $V\equiv Z_2\otimes Z'_2$, for instance, allows up to three stable particles, with two of them being unconditionally stable~\cite{Ma:2006uv,Cao:2007fy,Hur:2007ur}. All of them would, however, be real particles because  the group structure dictates that the terms $\phi_i^2$ are necessarily  allowed. In addition, the embedding of the Klein group into a gauge symmetry is non-trivial,  requiring the breaking chain $U(1)\otimes U'(1)\to Z_2\otimes Z'_2$ \cite{Petersen:2009ip} or a more elaborated one such as $SU(3)\to (SO(3)\to A_4)\to V$ \cite{Merle:2011vy}. Notice that the grand unification group $E_6$ yields up to two additional $U(1)$ factors when it is broken to the Standard Model~\cite{Langacker:2008yv}. For the case $Z_2\otimes Z'_2\otimes Z''_2$ up to eight DM particles may arise, with three being unconditionally stable and all of them being real. In constrast, the scenarios based on a single  $Z_N$ that we have studied predict that at most one of the scalar dark matter particles is real.

There are different ways in which one can go beyond the simplest scenarios considered in this work. One can imagine, for example, having not only scalars but also new fermions charged under the $Z_N$ \cite{Batell:2010bp}  and coupled among themselves via Yukawa interactions. Or one can replace the $Z_N$ by a better-motivated $U(1)$ local symmetry, as illustrated in appendix \ref{sec:uone}. Another possibility is to assume that the fields $\phi_i$ transform non-trivially under the SM gauge group. Two general scenarios arise in this case: $i$) all the fields $\phi$'s share the same SM quantum numbers and $ii)$ the DM particles transform under different $SU(2)_L$ representations\footnote{All the $\phi_i$'s are assumed to be color singlets.}. In both instances,  the scalar potential is similar to that for  SM singlets, but an important restriction arises from the fact that $\phi_i$ has to include a neutral particle, which is ensured by $Y=-2T_3$. Since direct detection searches exclude those dark matter candidates  having a direct coupling to the $Z$ boson (due to a spin independent cross section orders of magnitude larger than current bounds),  the possible values for the hypercharge that allow for a neutral particle  reduce to $Y=0$, which implies only $SU(2)_L$ representations of odd dimensionality. This means that only (complex) scalar fields transforming as a triplet, quintuplet or a septuplet with $Y=0$ are allowed to be part of the multicomponent DM scenario we are considering\footnote{The list of scalar $SU(2)_L$ multiplets as DM is finite once perturbativity of gauge couplings is imposed~\cite{Cirelli:2005uq}.}. The case of scalar doublet $\eta$ deserves a separate comment\footnote{See Ref. \cite{Ivanov:2012hc} for a $Z_N$-invariant DM scenario with several scalar doublets.}: since the term $\lambda_5(\eta^\dagger H)+$h.c. is forbidden for $Z_N$ with $N\geq3$, there is no mass splitting between the CP even and CP odd components of the neutral part. Therefore, inelastic scattering off nuclei is present, leading to a DD cross section ruled out by experiments. A detailed study of these possible extensions must, however, be left for future work.

\section{Conclusions}
\label{sec:conc}

We considered extensions of the Standard Model  by a number of  scalar fields that are SM singlets but have different charges under a new $Z_N$ ($N\geq 4$) symmetry and showed that they naturally lead to multi-component dark matter. We systematically analyzed these scenarios for $N\leq 10$ and for different sets of scalar fields. For $N$ odd, the dark matter particles turned out to be  complex scalar fields  whereas one of them may be a real scalar field for $N$ even. The regions of the parameter space where multi-component dark matter can be realized were determined analytically and illustrated graphically for  up to five dark matter particles.  Usually, these regions  depend on the masses of the scalar fields, but in some special cases we found \emph{unconditional} stability. A common feature of these scenarios is the appearance of multiple dark matter conversion processes as well as semi-annihilations. Many new models for multi-component dark matter can be implemented within this simple setup. 

\section*{Acknowledgments}
Work supported by Sostenibilidad-UdeA and the UdeA/CODI Grant
2017-16286, and by COLCIENCIAS through the Grant 111577657253. 
O.Z. acknowledges the ICTP Simons associates program. 

\appendix
\section{Unconditional stability}
\label{sec:unco}
Let's recall that for $N$ prime only one particle can be stable by symmetry reasons (the particle having a non trivial $Z_N$ charge in the bottom of the mass spectrum). For instance, $Z_5$ and $Z_7$.  
On the other hand, for $p,q$ coprimes then 
\begin{align}
    Z_N\cong Z_p\otimes Z_q.
\end{align}
As some concrete examples, $Z_6\cong Z_2\otimes Z_3$ and $Z_{10}\cong Z_2\otimes Z_5$. Thus in principle there may be two stable particles.
Under $Z_N$, $Z_p$ and $Z_q$ symmetries we have the following charges:
\begin{align}
    Z_N:&\,\,{1,w_N,w_N^2,...,w_N^{N-1}}, \,\,\,\text{with}\,\,\,w_N=e^{i2\pi/N};\\
    Z_p:&\,\,{1,w_p,w_p^2,...,w_p^{p-1}}, \,\,\,\text{with}\,\,\,w_p=e^{i2\pi/p};\\
    Z_q:&\,\,{1,w_q,w_q^2,...,w_q^{q-1}}, \,\,\,\text{with}\,\,\,w_q=e^{i2\pi/q}. 
\end{align}
A field transforming under $Z_p\otimes Z_q$ has the following charge:
\begin{align}
    e^{i2\pi n_p/p}e^{i2\pi n_q/q}=e^{\frac{i2\pi}{pq}(qn_p+pn_q)}=w_N^{(qn_p+pn_q)}=w_N^{n_N},
\end{align}
where $n_N=qn_p+pn_q$, $n_p=0,1,..,p-1$, $n_q=0,1,..,q-1$ and $n_N=0,1,..,N-1$. Thus a field transforming under $Z_p\otimes Z_q$ as $w_p^{n_p}\times w_q^{n_q}$, has a charge under $Z_{pq}=Z_N$ equals to $w_N^{n_N}=w_N^{(qn_p+pn_q)}$. It follows that the two stable particles, one associated to $Z_p$ and the other one associated to $Z_q$ must transform trivially under the other symmetry. 
Hence, $\phi$ is singlet under $Z_p$ if $n_p=0$, which implies a $Z_N$ charge $w_N^{n_N}=w_N^{pn_q}$. Since $n_N$ is integer, then the possible charges for $\phi$ under $Z_q$ are those satisfying $n_q=n_N/p\in\{1,2,...,q-1\}$. 
In the same form, if $\chi$ is a singlet under $Z_q$ the possible charges under $Z_p$ are those satisfying $n_p=n_N/q\in\{1,2,...,p-1\}$. 
For ilustration purposes we consider two examples:
\begin{itemize}
    \item $Z_6\cong Z_2\otimes Z_3$:   
    \begin{align}   
    \phi\sim&[(1,w_3)\lor(1,w_3^2)]\,\,\text{under}\,\,(Z_2,Z_3);\,\,\, \phi\sim [w_6^2\lor w_6^4]\,\,\text{under}\,\,Z_6.\\
    \chi\sim&(w_2,1) \,\,\text{under}\,\,(Z_2,Z_3);\,\,\,\chi\sim w_6^3\,\,\text{under}\,\,Z_6.
\end{align}
Therefore $(\phi_2,\phi_3)$ is the unique possible scenario with two stable fields under $Z_6$. 
\item $Z_{10}\cong Z_2\otimes Z_5$:   
    \begin{align}   
    \phi\sim&[(1,w_5)\lor(1,w_5^2)\lor(1,w_5^3)\lor(1,w_5^4)]\,\,\text{under}\,\,(Z_2,Z_5);\nonumber\\
    &\,\,\, \phi\sim [w_{10}^2\lor w_{10}^4\lor w_{10}^6\lor w_{10}^8]\,\,\text{under}\,\,Z_{10}.\\
    \chi\sim&(-1,1) \,\,\text{under}\,\,(Z_2,Z_5);\chi\sim w_{10}^5\,\,\text{under}\,\,Z_{10}.
\end{align}
Therefore $(\phi_2,\phi_5)$ and $(\phi_4,\phi_5)$ are the unique possible scenarios with two stable fields under $Z_{10}$. 
\end{itemize}
Following this reasoning, the simplest scenario featuring unconditional stability for three fields is realized via a $Z_{30}\cong Z_2\otimes Z_3\otimes Z_5$ symmetry. However, unconditional stability at the renormalizable level for three particles can be first obtained with a  $Z_{18}$ symmetry and any of the following sets of fields $(\phi_1,\phi_6,\phi_9)$, $(\phi_5,\phi_6,\phi_9)$, or $(\phi_6,\phi_7,\phi_9)$.

\section{$U(1)$ completion}
\label{sec:uone}

One of the advantages of using a $Z_N$ symmetry is that such a setup can be easily embedded within extensions of the SM including an extra $U(1)_X$ gauge symmetry, as we now illustrate with an example.  Let us embed the scenario with  a $Z_8$ and  fields $(\phi_1,\phi_2,\phi_3)$ previously discussed in section \ref{sec:complex}. We would then replace the $Z_8$ with a  $U(1)_X$ local symmetry under which the  charges of the three fields are respectively $1,2,3$. This gauge symmetry is assumed to be  spontaneously broken by the vacuum expectation value of a SM singlet scalar $S$ with $X$-charge equal to $8$. The most general $U(1)_X$-invariant scalar potential is then given by 
\begin{align}\label{eq:U1Xphi123}
 \mathcal{V}&=\,\,\mathcal{V}(H,S)+\sum_{i=1}^3\left[\mu_{\phi_i}^2|\phi_i|^2+\lambda_{\phi_i}|\phi_i|^4+\lambda_{H\phi_i}|H|^2|\phi_i|^2+\lambda_{S\phi_i}|S|^2|\phi_i|^2\right]+\sum_{i<j}\lambda_{\phi_i\phi_j}|\phi_i|^2|\phi_j|^2\nonumber\\
 & + \left[\kappa_1\phi _2 \phi _1^{*2} + \kappa_2\phi_3\phi_1^*\phi_2^* +\lambda_1\phi_2\phi_3^2S^* + \lambda_2\phi_3\phi_1^{*3} + \lambda_{3}\phi_2^2\phi_1^*\phi_3^* + \text{H.c.}\right].
 \end{align}
Here $\mathcal{V}(H,S)=-\mu^2_H|H|^2+\lambda_H|H|^4-\mu^2_S|S|^2+\lambda_S|S|^4+\lambda_{SH}|S|^2|H|^2$ is the scalar potential involving only the SM Higgs and $S$ fields.  From that potential, one can see that the three fields will be stable if the condition $M_2<2 M_1,\,M_1<M_2+M_3,\, M_3<M_1+M_2,\, M_2<2 M_3$, %$\left(\frac{M_1}{3}<M_3\leq M_1\land M_1-M_3<M_2<2 M_3\right)\lor\left(M_1<M_3<3 M_1\land M_3-M_1<M_2<2 M_1\right)$
is fulfilled, which is the same stability condition found for the scenario with the  fields  $(\phi_1,\phi_2,\phi_3)$ and $Z_8$ invariance. In the local model, the new scalar fields have additional interactions mediated by the  $U(1)_X$ gauge boson but they do not bring about new decay processes.  Notice that the $U(1)_X$ invariance is more restrictive than the $Z_8$ as it allows only the quartic terms that have a total $Z_8$ charge equal to zero --see Eq.~(\ref{eq:Z8phi123-L4}).  
   
   In this way, we can reproduce most of the features of a $Z_N$ model but using  a local (gauge) symmetry rather than a discrete one. 
   Moreover, the $U(1)$ origin of the $Z_N$ stabilizing symmetry may be used to relate multi-component DM scenarios with the solution to small-scale structure problems \cite{Bernal:2015bla,Choi:2015bya} or to the open questions of the SM such as the origin of the neutrino masses \cite{Heeck:2012bz,Bernal:2018aon}, the flavor puzzle \cite{Sierra:2014kua} and the CP violation in strong interactions \cite{Carvajal:2018ohk}, 
% \cite{Heeck:2012bz} for a three Dirac fermions scenario with a left over $Z_{11}$ symmetry.
% \cite{Bernal:2018aon} for a fermion and a scalar scenario with a left over $Z_{7}$ symmetry.
\bibliographystyle{apsrev4-1long}
\bibliography{references}

\end{document}